\documentclass[aps,prx,twocolumn,amsmath,amssymb,reprint,numbers,superscriptaddress,noeprint,longbibliography]{revtex4-2}

\usepackage{braket}
\usepackage{geometry}
\usepackage{graphicx}
\usepackage{color}
\usepackage{caption}
\usepackage{textcomp}
\usepackage{xcolor}
\usepackage[normalem]{ulem}
\usepackage{style}
\hypersetup{
     colorlinks   = true,
     linkcolor    = blue,
     citecolor    = blue,
     urlcolor = blue
}

\newcommand{\up}{$\ket{\uparrow}$}
\newcommand{\down}{$\ket{\downarrow}$}
\newcommand{\Do}{D\textsuperscript{0}}
\newcommand{\DoX}{D\textsuperscript{0}X}
\newcommand{\InZnX}{In$_\mathrm{Zn}^\mathrm{0}$X}

\newcommand{\AlX}{Al$^\mathrm{0}$X}
\newcommand{\GaX}{Ga$^\mathrm{0}$X}
\newcommand{\InX}{In$^\mathrm{0}$X}
\newcommand{\Ino}{In\textsuperscript{0}}
\newcommand{\InIon}{In$^\mathrm{+}$}
\newcommand{\InZn}{In$_\mathrm{Zn}^\mathrm{0}$}
\newcommand{\transInX}{\InZn\,$\leftrightarrow$\,\InZnX}
\newcommand{\sigplus}{$\sigma^+$}
\newcommand{\sigminus}{$\sigma^-$}
\newcommand{\Otwo}{O\textsubscript{2}}

\definecolor{cola}{rgb}{0.7,0.1,0.1}

\definecolor{pink}{rgb}{1, 0.412, 0.702}
\definecolor{colb}{rgb}{0.9,0.4,0}

\begin{document}

\title{Coherent Microwave Control of Optically Addressable Donor Qubits in ZnO}

\author{Ethan R. Hansen}
\affiliation{Department of Physics, University of Washington, Seattle, WA, 98195, USA}
\author{Dong-Rong Wu}
\affiliation{Department of Chemistry, University of Wisconsin-Madison, 53706, USA}
\author{Yixuan Li}
\affiliation{Department of Electrical and Computer Engineering, University of Washington, Seattle, WA, 98195, USA}
\author{Yaser Silani}
\affiliation{Department of Physics, University of Washington, Seattle, WA, 98195, USA}
\author{Joseph~Falson}
\affiliation{Department of Applied Physics and Materials Science, California Institute of Technology, Pasadena, California 91125, USA.}
\affiliation{Institute for Quantum Information and Matter, California Institute of Technology, Pasadena, California 91125, USA.}
\author{Yusuke Kozuka}
\affiliation{Department of Applied Physics and Quantum-Phase Electronics Center (QPEC), University of Tokyo, 113-8656 Tokyo, Japan}
\author{Masashi Kawasaki}
\affiliation{Department of Applied Physics and Quantum-Phase Electronics Center (QPEC), University of Tokyo, 113-8656 Tokyo, Japan}
\affiliation{Center for Emergent Matter Science (CEMS), RIKEN, 351-0198 Saitama, Japan}
\author{Yuan Ping} 
\affiliation{Department of Materials Science and Engineering, University of Wisconsin-Madison, 53706, USA}
\affiliation{Department of Physics, University of Wisconsin-Madison, 53706, USA}
\affiliation{Department of Chemistry, University of Wisconsin-Madison, 53706, USA}
\author{Kai-Mei C. Fu}
\affiliation{Department of Physics, University of Washington, Seattle, WA, 98195, USA}
\affiliation{Department of Electrical and Computer Engineering, University of Washington, Seattle, WA, 98195, USA}
\affiliation{Physical Sciences Division, Pacific Northwest National Laboratory, Richland, WA, 99352, USA}

\begin{abstract}

Optically addressable shallow donors in ZnO combine efficient spin-selective optical transitions with the potential for long spin coherence in an isotopically purifiable host lattice, making them an attractive platform for spin-photon quantum technologies. A key missing capability, however, has been coherent control beyond the small-angle rotations accessible with ultrafast optical pulses. Here we demonstrate coherent microwave control of implanted $^{115}\mathrm{In}$ donors in ZnO. Resonant optical pumping initializes and reads out the donor electron spin. Pulsed optically-detected magnetic resonance resolves the ten hyperfine transitions associated with the coupled $^{115}\mathrm{In}$ nuclear spin (I = 9/2) and reveals optical-pumping-induced nuclear spin polarization. We observe coherent Rabi oscillations with a maximum Rabi frequency of $\Omega/2\pi = 36.2 \pm 0.7$\;MHz, corresponding to a $\pi$-pulse time of 13.8$\pm$0.3\;ns, and characterize the spin coherence using Ramsey, Hahn echo and dynamical-decoupling measurements. Unexpectedly, the measured coherence is substantially shorter than reported in previous optical studies of donor spins in ZnO at high magnetic field. Control experiments rule out several simple explanations including microwave heating and instantaneous diffusion from the driven donor ensemble, leaving an open question regarding the origin of decoherence at low magnetic field in microwave-controlled ZnO donors. These results establish microwave control of ZnO donor qubits with resonant optical access to specific donor species. More broadly, they demonstrate that coherent microwave control can be achieved in optically addressable spin systems with nanosecond-scale inhomogeneous dephasing, opening a route to field-, temperature-, and materials-dependent investigation of coherence-limiting mechanisms, studies inaccessible to all-optical control methods, and to the development of optically interfaced electron–nuclear spin registers in emerging donor and defect-spin platforms.

\end{abstract}

\date{\today}

\maketitle

\section{Introduction}

Optically accessible solid-state defects are promising platforms for quantum information because they combine long-lived spin states with optical interfaces for spin-photon quantum technologies~\cite{benjamin2009qce,weber2010qcd,ladd2010qc,wehner2018qiv,orieux2016rai}.
Zinc oxide (ZnO), a II-VI semiconductor, is an attractive host for optically addressable spin qubits. Its direct band gap enables efficient optical transitions, while its weak spin-orbit coupling~\cite{niaouris2022esr} and potential for an isotope-purified, low-nuclear-spin host lattice are favorable for long spin coherence, as demonstrated in platforms such as silicon and diamond~\cite{tyryshkin2012esc,balasubramanian2009usc}. Previous studies of neutral shallow donors in ZnO--including Al, Ga, and In substituting on the Zn site--have reported donor-bound-exciton transitions narrow enough for optical spin manipulation~\cite{wagner2011bez,linpeng2018cps,wang2023pdq,niaouris2023col}, electron spin lifetimes up to 0.5\,s~\cite{niaouris2022esr}, and coherence times up to 50~\textmu s in high-purity substrates~\cite{linpeng2018cps}. Indium donors are particularly appealing because the donor electron spin is strongly coupled to the $^{115}$In nuclear spin ($I=9/2$) through an approximately 100\,MHz hyperfine interaction~\cite{block1982odm,gonzalez1982mrs,buss2016omm}, providing access to a long-lived nuclear-spin memory~\cite{neumann2008mea,morton2008ssq,merkel2008qch} and a coupled electron-nuclear spin register. Together with the recent demonstration of implanted In donor ensembles~\cite{wang2023pdq} and isolated single In donors in ZnO~\cite{hansen2024isd}, these results establish In donors in ZnO as a promising platform for optically addressable spin qubits and electron-nuclear spin registers.

A remaining challenge is coherent control of the donor electron spin with rotations large enough for general qubit manipulation, specifically $\pi$-pulses on timescales shorter than the inhomogeneous dephasing time $T_2^*$, which requires resonant microwave fields at millitesla scale. Previous coherent control of shallow donors in ZnO was achieved using ultrafast optical pulses; however, this approach was limited to small-angle rotations, with the ensemble population transfer saturating near 40\% with increasing laser power because of laser-induced dephasing~\cite{linpeng2020dqd}. Direct microwave (MW) control provides a complementary route which avoids optical dephasing, enables full spin rotations, and provides high-resolution access to the hyperfine-resolved electron-nuclear spin manifold. Integrating microwave control with resonant optical initialization and readout combines coherent spin manipulation with donor-specific optical addressing, a capability that is central to donor-based quantum technologies. 

Microwave control of solid-state defect qubits is well established in platforms such as nitrogen-vacancy and group-IV vacancy centers in diamond~\cite{karapatzakis2024mct} and point defects in silicon carbide~\cite{koehl2011rtc}. However, realizing coherent control is particularly challenging in the donor:ZnO system due to the short inhomogeneous dephasing time ($T_2^* \approx 17$\,ns)~\cite{linpeng2020dqd}, which necessitates millitesla-scale resonant microwave fields to drive spin rotations within the coherence window. In addition, the Faraday optical geometry used for spin-selective initialization and readout requires the microwave magnetic field to be perpendicular to the static magnetic field and therefore parallel to the ZnO surface. While large microwave magnetic fields are readily generated in geometries with fields oriented normal to a crystal surface, achieving comparable field strengths parallel to the surface while preserving optical access is substantially more challenging.

In this work, we demonstrate coherent microwave control of implanted In donors in ZnO using a microwave resonator geometry that combines strong in-plane microwave driving fields with optical access. This architecture enables resonant optical initialization and readout while providing the millitesla-scale microwave fields needed for fast spin rotations. We use resonant excitation to initialize and read out implanted In donor electron spins, and then perform pulsed ODMR to resolve the ten hyperfine transitions arising from the coupling of the electrons to the $^{115}$In nuclear spin ($I=9/2$), revealing optical-pumping-induced nuclear spin polarization. 
We then demonstrate full coherent rotations of the donor electron spin through Rabi oscillations and characterize the spin coherence using Ramsey, Hahn-echo, and CPMG pulse sequences. Unexpectedly, the measured Hahn-echo is substantially shorter than reported in previous optical studies of donor spins in ZnO at high magnetic field, and only modestly improves with increasing the number of refocusing pulses. A series of control experiments point to magnetic noise from the local spin environment as the likely source of the shortened coherence which must be understood and controlled to access the full potential of an isotopically-purified ZnO donor qubit host. 


\section{Spin System and Optical Properties}

For this study, we utilize a 300\,\textmu m-thick ZnO crystal (Tokyo Denpa) with a 3\,\textmu m-thick epilayer grown via molecular beam epitaxy on the (0001) surface. 
The epilayer is masked with a square TEM grid and implanted with In ions at a concentration of $10^{11}$\,ions/$\mathrm{cm}^2$ and an energy of $380$\,keV. The sample is annealed at 700\,$^\circ$C in an oxygen atmosphere for one hour. Simulation of the implantation~\cite{ZieglerSRIM} yields a peak implantation depth of $90$\,nm with a peak density of $10^{16}$\,ions/$\mathrm{cm}^3$ (App.~\ref{app:samp_prep}).

Experiments are performed in a helium immersion cryostat (gas or liquid) with a split-coil superconducting magnet and base temperature of $1.7$\,K. Optical characterization is performed with a custom confocal microscope with sub-micron spatial resolution (App.~\ref{app:experiment}).

\begin{figure}[ht]
    \centering
    \includegraphics[width=0.9\linewidth]{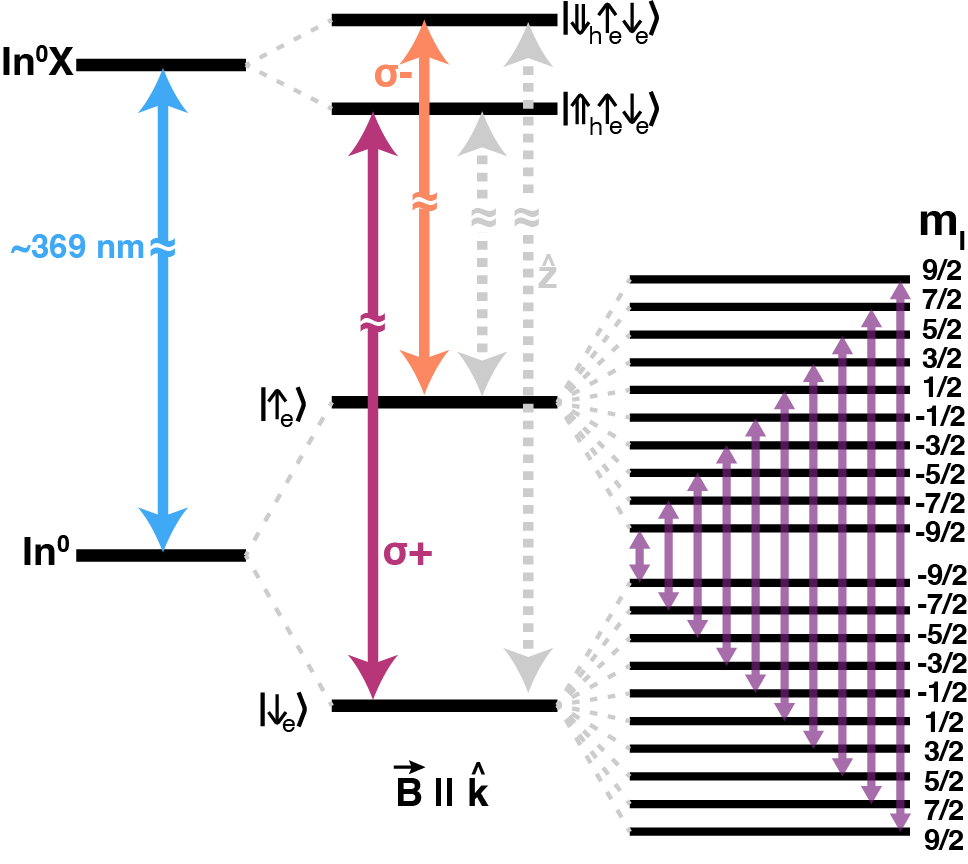}
    \caption{
    Energy-level structure of a neutral indium donor in ZnO in Faraday geometry, $\vec{B}\parallel\hat{k}\parallel\hat{c}$, in which $\hat k$ is the optical axis and $\hat c$ is the ZnO c-axis. The neutral donor ground state (\Ino) consists of two electron-spin states split by the Zeeman interaction. Optical excitation near $369$\,nm drives transitions to the donor-bound exciton state (\InX), whose two hole-spin states are also Zeeman split. The spin-selective optical transitions are labeled by their circular polarization, $\sigma^+$ and $\sigma^-$. Hyperfine coupling to the $^{115}$In nuclear spin ($I=9/2$) further splits each electron-spin manifold into ten sublevels labeled by $m_I$.
    }
    \label{fig:spinsystem}
\end{figure}

The energy-level structure of the shallow indium donor is shown in Fig.~\ref{fig:spinsystem} in the Faraday geometry. In an external magnetic field $B_{\mathrm{ext}}$, the two \Ino\ electron spin states are split by the Zeeman interaction, with energy splitting $\Delta E_e = g_e \mu_B B_{\mathrm{ext}}$, where $g_e$ is the electron $g$-factor and $\mu_B$ is the Bohr magneton. The \InX\ state is split by the hole Zeeman interaction, with $\Delta E_h = g_h \mu_B B_{\mathrm{ext}}$, where $g_h$ is the hole $g$-factor. The difference between $g_h$ and $g_e$ enables electron spin-selective optical transitions at high field. These spin-selective optical transitions provide the mechanism for resonant initialization and readout of the donor electron spin used in the microwave control experiments below. Hyperfine coupling to the $^{115}$In nuclear spin ($I=9/2$) further splits each electron spin manifold into ten levels with an approximately 50\,MHz spacing~\cite{block1982odm}.

\begin{figure}[ht]
    \centering
    \includegraphics[width=0.9\linewidth]{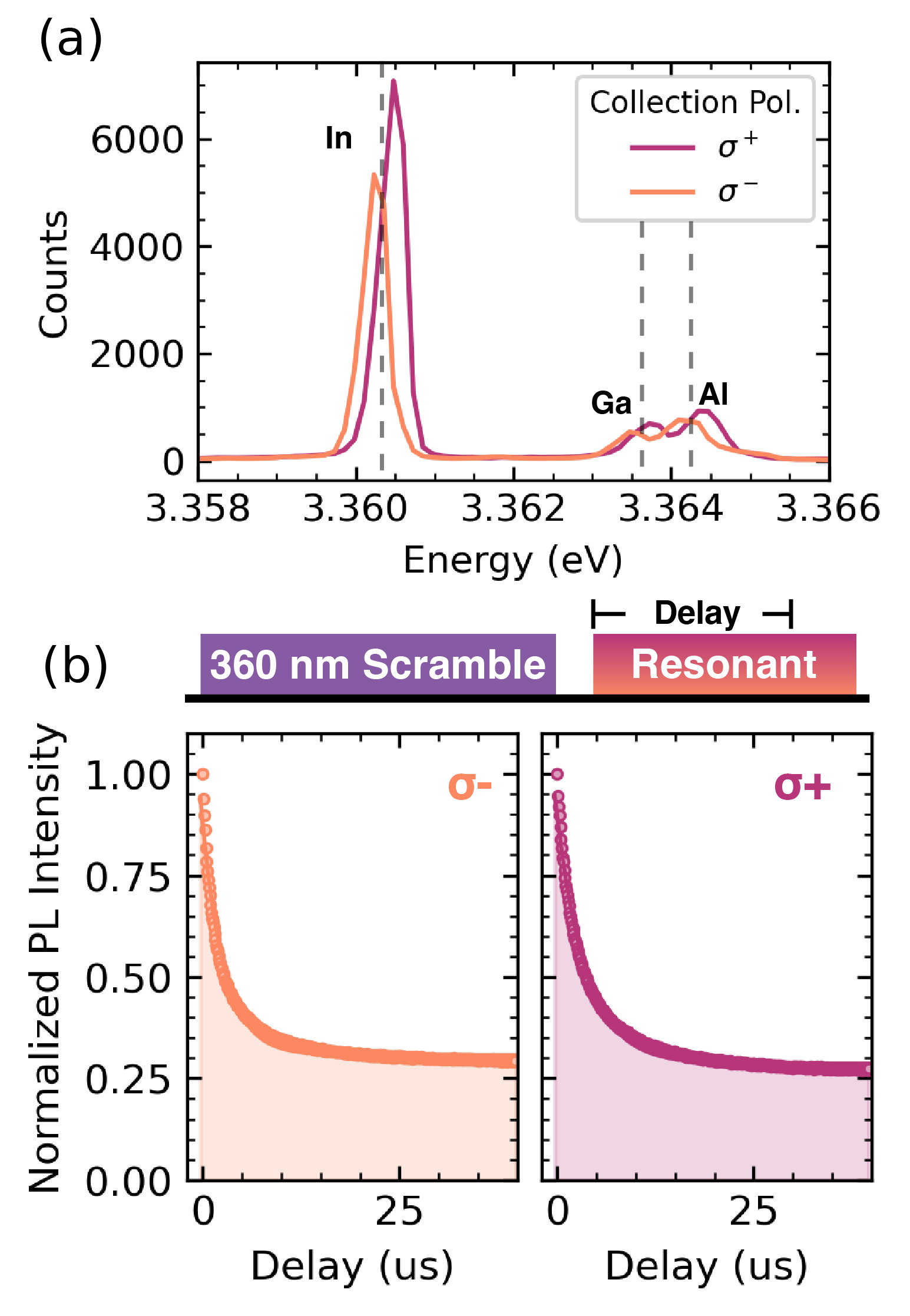}
    \caption{
    (a) \InX\ photoluminescence from implanted In at $B=6$\,T collected in \sigplus\ and \sigminus\ polarization. Also observed are Ga and Al transitions originating from the substrate. Dashed lines label the zero-field donor-bound-exciton transitions for In, Ga, and Al donors. Spectra are instrument resolution limited. Excitation wavelength is $360$\,nm
    (b) Time-resolved optical pumping at $B=300$\,mT following a $360$\,nm scramble pulse and resonant excitation on the \sigminus\ ($369.26781$\,nm) and \sigplus\ ($369.26457$\,nm) transitions. The scramble and resonant pump powers are each $150$\,nW and applied for a duration of $150$\,\textmu s. Traces are background-subtracted and normalized.
    T = $5.5$\,K.
    }
    \label{fig:opticalAccess}
\end{figure}

Fig.~\ref{fig:opticalAccess} depicts the polarization-dependent photoluminescence spectrum under above band excitation at 6\,T in which both frequency and polarization selectivity (due to Faraday geometry) are visible. Microwave experiments, however, are performed at $B_{\mathrm{ext}}=300$\,mT, corresponding to an electron spin splitting of approximately $8.2$\,GHz. At this field, the electron Zeeman splitting is comparable to the optical linewidth of $8.44$\,GHz (App.~\ref{app:optical_characterization}), so the two spin-selective resonant pump transitions are not fully resolved. As a result, spin initialization and readout must also rely on polarization selectivity, requiring the Faraday geometry. 

 We demonstrate spin-selective initialization in Fig.~\ref{fig:opticalAccess}(b). To reset the spin population, we first apply an above-band 360\,nm pulse. We then resonantly drive the \sigminus\ (\sigplus) transition (Fig.~\ref{fig:spinsystem}) to pump into \down\ (\up). We record photoluminescence as a function of time during the resonant excitation to track the optical pumping dynamics and resulting spin contrast. Assuming the above-band excitation pulse prepares an equal population in the two spin states, the measured optical pumping contrasts of $73\%$ for \sigplus\ excitation and $71\%$ for \sigminus\ excitation correspond to spin initialization fidelities of $86\%$ into \up\ and $85\%$ into \down, respectively. The spin initialization fidelity is likely limited by imperfect spectral and polarization selectivity, which allow the resonant pump laser to partially drive the opposite spin transition and reduce the optical pumping contrast. The resulting spin visibility is sufficient for high-contrast initialization and readout in the microwave-control experiments presented below.


\section{Microwave Control}
\label{sec:spin_control}

\subsection{Strong Microwave Driving in Faraday Geometry}

Coherent microwave control in natural-abundance ZnO requires mT-scale microwave magnetic fields to drive spin rotations on timescales shorter than the inhomogeneous dephasing time ($T_2^* \approx 17$\;ns), while maintaining optical access in the Faraday geometry. Achieving such fields is challenging because the microwave magnetic field must be oriented parallel to the ZnO surface and perpendicular to the optical axis.  Conventional planar microwave structures  are difficult to use because placing the sample on top of a planar microwave device would require optical access through the back side of the crystal. This is inefficient for resonant donor-bound-exciton excitation and collection due to the large optical depth of the ZnO substrate at the donor-bound-exciton transition~\cite{niaouris2023col}.

\begin{figure}[ht]
    \centering
    \includegraphics[width=0.9\linewidth]{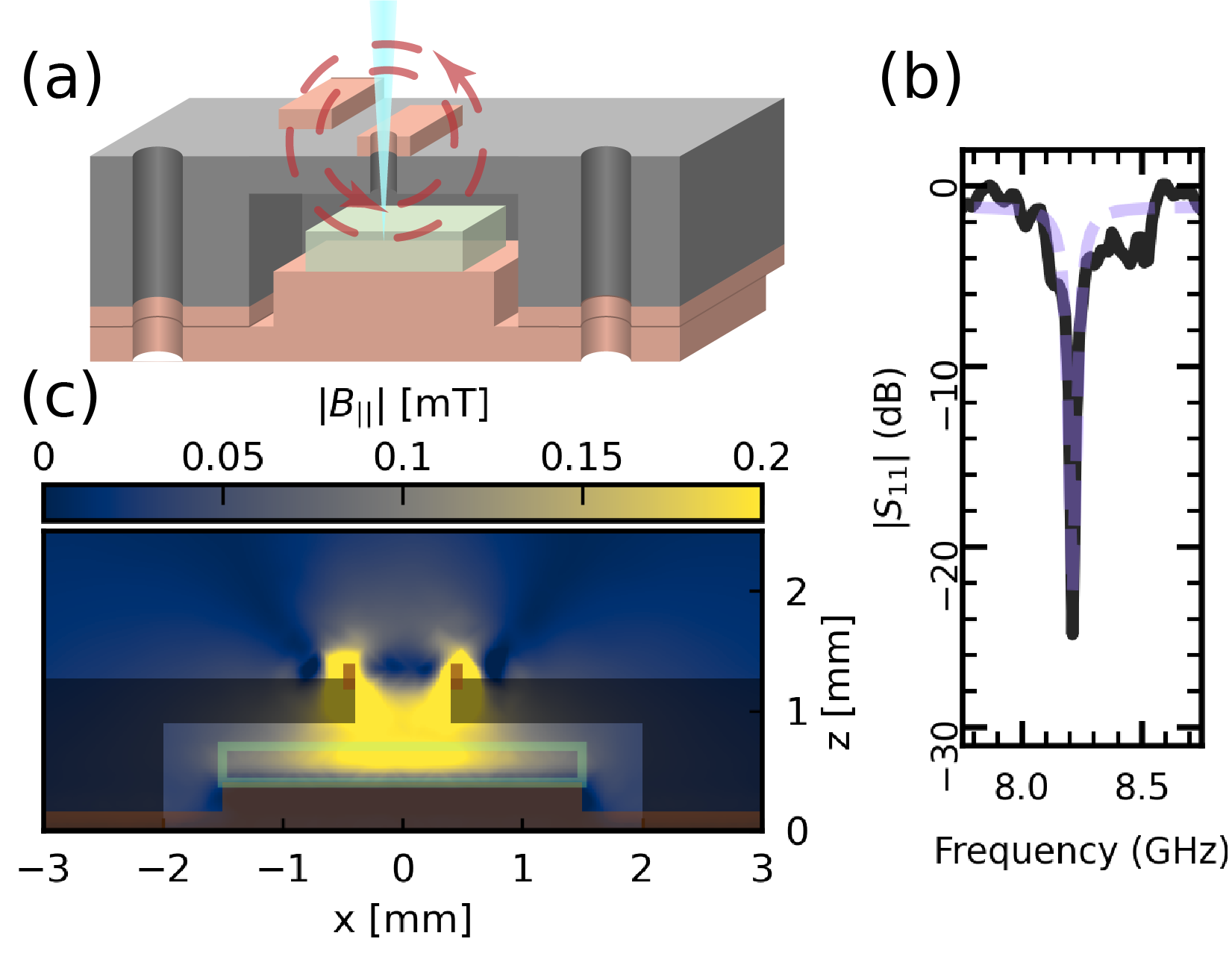}
    \caption{
    Microwave resonator for coherent spin control.
    (a) Schematic of the microstrip resonator with the embedded ZnO sample. The blue beam indicates optical access through the central hole, and the red arrows indicate the microwave magnetic field.
    (b) Measured reflection spectrum, $|S_{11}|$, showing a resonance near $8.2$\,GHz. Dashed purple curve is the simulated $|S_{11}|$.
    (c) Simulated in-plane microwave magnetic field, $|B_{\parallel}|$, for an applied microwave power of $0.5$\,W. (CST Microwave Studio~\cite{CST2024}.)
    }
    \label{fig:resonator}
\end{figure}

We therefore embed the ZnO sample within a microstrip resonator, as shown in Fig.~\ref{fig:resonator}(a). This geometry places the implanted donor ensemble in a region of strong microwave magnetic field while preserving optical access through a 0.8\,mm diameter aperture at the center of the device. The resulting microwave magnetic field is predominantly parallel to the ZnO surface and therefore perpendicular to the external magnetic field in the Faraday geometry. The resonator was designed for high-power pulsed operation, with peak microwave powers of up to 100\,W.

Figure~\ref{fig:resonator}(b) shows the measured reflection spectrum of the resonator, which exhibits a resonance near 8.2\,GHz with a full width at half maximum of approximately 44\,MHz. Finite-element simulations of the resonator field distribution are shown in Fig.~\ref{fig:resonator}(c). At a depth of 100\,nm below the ZnO surface, corresponding approximately to the peak of the implanted donor distribution, the simulated in-plane microwave magnetic field amplitude is 0.215\,mT for an applied microwave power of 0.5\,W. The corresponding Rabi frequency is given by $ \Omega_R = g_e\mu_B B_{MW}/\hbar$, where $B_{\mathrm{MW}}$ is the microwave magnetic field amplitude. Scaling this field amplitude to the maximum applied power of $100$\,W yields $B_{\mathrm{MW}} \approx 3$\,mT, corresponding to an ideal Rabi frequency of $\Omega_R/2\pi \approx 83$\,MHz and $t_\pi \approx 6$\,ns. These field strengths are sufficient to perform coherent spin rotations within the $T_2^*$ window of ZnO donors and motivate the use of this resonator geometry for microwave control in the Faraday configuration. 

\subsection{Hyperfine-Resolved Pulsed ODMR}
\begin{figure}[ht]
    \centering
    \includegraphics[width=0.8\linewidth]{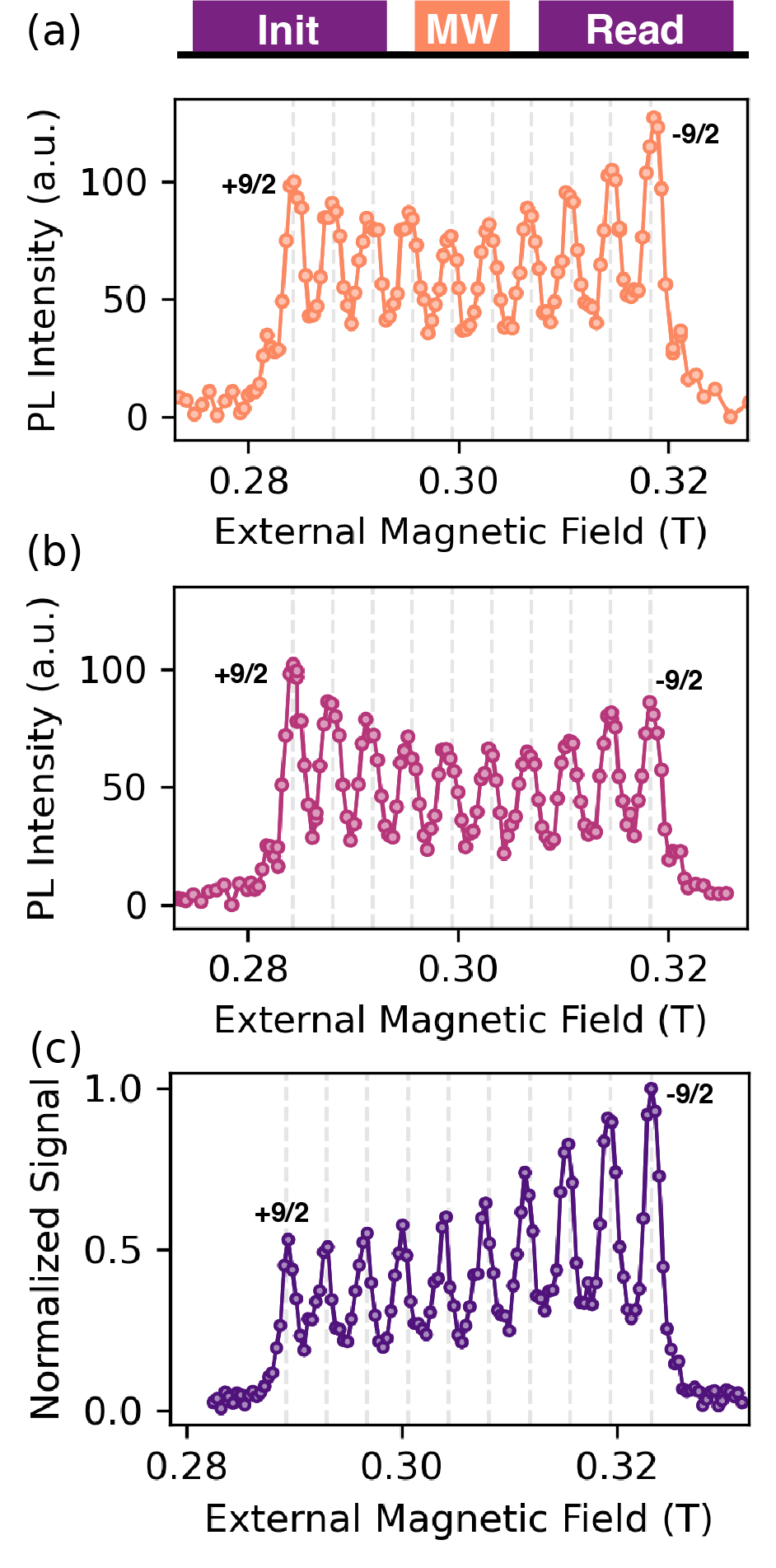}
    \caption{
    (a,b) Pulsed ODMR spectra acquired by sweeping the external magnetic field at fixed microwave frequency ($f_{\mathrm{MW}}=8.20$\,GHz) with resonant optical pumping on the \sigminus\ and \sigplus\ transitions, respectively. 
    $\Omega_R/2\pi = 23$\,MHz for panels (a) and (b). 
    (c) Pulsed ODMR spectrum acquired with \sigminus\ pumping after tilting the sample by approximately $13^\circ$ relative to the configuration used in panels (a) and (b); $\Omega_R/2\pi = 19$\,MHz and $f_{\mathrm{MW}}=8.33$\,GHz.
    Vertical dashed lines are split by $104$\,MHz and aligned with the 10 hyperfine transitions.
    T=5.5\,K.
    }
    \label{fig:odmr}
\end{figure}

We first use pulsed optically detected magnetic resonance (PODMR) to establish hyperfine-resolved microwave control of the implanted In donor ensemble. Earlier ODMR, ENDOR, and ODENDOR studies of ZnO donors relied on thermal polarization and electrical~\cite{hofmann2002hrs} or indirect optical detection via donor-acceptor pair recombination~\cite{block1982odm,gonzalez1982mrs}. In contrast, resonant excitation of the implanted In donor-bound-exciton transition enables deterministic electron-spin initialization and direct optical readout of microwave-driven spin transfer. In this measurement, the external magnetic field is swept while the microwave frequency is held fixed near the resonator frequency. We resonantly initialize the donors into $|\uparrow\rangle$ ($|\downarrow\rangle$) using the $\sigma^+$ ($\sigma^-$) transition, apply a microwave pulse, and then read out the spin population using a second resonant pulse on the same optical transition.

The resulting PODMR spectra are shown in Fig.~\ref{fig:odmr}. We observe ten resonances corresponding to the ten $^{115}$In hyperfine states with a splitting of $\sim104$\,MHz, in agreement with previous measurements~\cite{block1982odm,gonzalez1982mrs}. The hyperfine splitting exceeds the measured linewidth, allowing the ten nuclear-spin manifolds to be individually resolved and enabling selective addressing  of the electron-spin transition for each nuclear-spin state. The measured linewidths reflect the combined effects of the finite $\pi$-pulse excitation bandwidth and the inhomogeneous broadening corresponding to $T_2^* \approx 17$\,ns~(App.~\ref{app:odmr_broadening}). 

The relative peak amplitudes exhibit a pronounced asymmetry across the hyperfine manifold. The observed spectral asymmetry can be qualitatively explained by an Overhauser-type polarization mechanism. Under resonant optical pumping, electron spin polarization is transferred to the nuclear spin system through hyperfine interactions, leading to a non-equilibrium nuclear-spin population distribution and consequently asymmetric ODMR spectra~\cite{jeffries1960dnp}. The reversal of the asymmetry under excitation of the opposite spin-selective optical transition (Fig.~\ref{fig:odmr}b) is consistent with a reversal of the optically induced nuclear-spin polarization .

In addition to the polarization-dependent asymmetry, the spectra exhibit a bowl-shaped amplitude envelope with reduced intensity near the central hyperfine transitions. This feature depends on the alignment between the ZnO $\hat c$ axis and the external magnetic field. As shown in Fig.~\ref{fig:odmr}(c), rotating the sample with respect to the applied magnetic field suppresses the bowl shape while preserving the nuclear-spin polarization bias. The experimental results suggest eigenstate mixing with field alignment, with Fig.~\ref{fig:odmr}(c) corresponding to a nearly aligned field. However, the donor electron spin-1/2 ground state is highly isotropic in ZnO; nor can the bowl shape be attributed to the In electric quadrupole moment (App.~\ref{app:odmr_bowl-shape}). The physical origin of the anisotropy, either directly reflecting the nuclear spin polarization or readout/initialization effects via the anisotropic excitonic state, remains an open question and will be investigated in future work. 

Taking the transition intensity to correspond to nuclear spin polarization in Fig.~\ref{fig:odmr}(c), we obtain a nuclear spin polarization $P_N = \langle I_z\rangle/I = -0.19$. While the extracted nuclear spin polarization is modest, it is obtained during the PODMR measurement conditions and is not optimized for nuclear spin polarization. The observation of hyperfine-resolved microwave control together with optically-induced nuclear-spin polarization establishes hyperfine-resolved access to the coupled electron-nuclear spin system and provides a foundation for future studies of electron-nuclear spin interactions, control, and nuclear spin coherence.   

\subsection{Coherent Spin Rotations}

\begin{figure}[ht]
    \centering
    \includegraphics[width=0.95\linewidth]{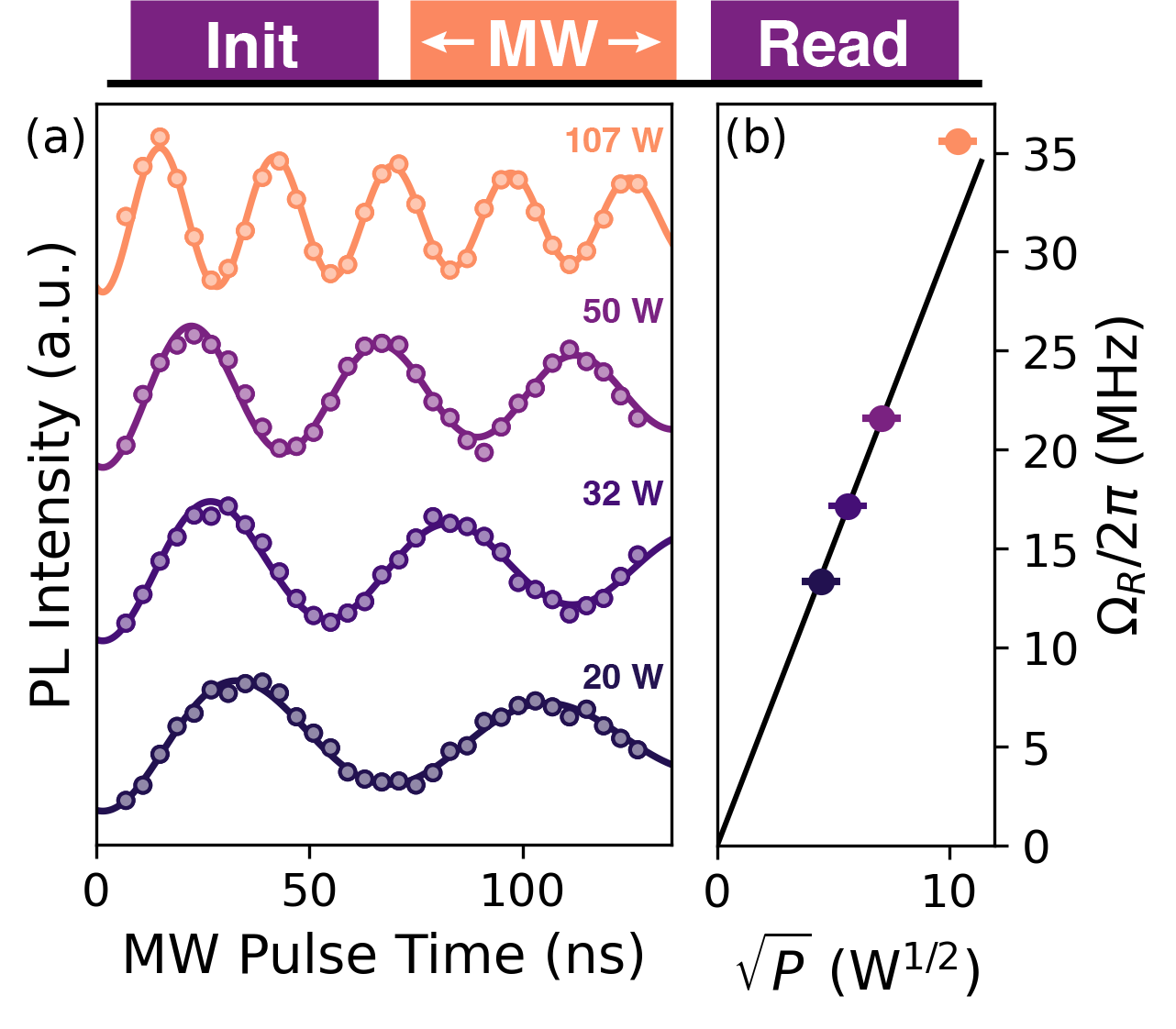}
    \caption{
    (a) Pulsed Rabi measurements on the $m_I=-9/2$ hyperfine transition for four applied microwave powers. Acquired with \sigminus\ optical pumping.
    The traces are vertically offset for clarity. Solid curves are a simultaneous fit to an ensemble Rabi model in which the coherent two-level response is averaged over a Gaussian distribution of static microwave detunings with fixed width $\sigma_\Delta/2\pi=13$\,MHz, corresponding to $T_2^*\simeq17$\,ns for a Gaussian Ramsey envelope.
    No additional homogeneous decay is included in these curves, i.e. $T_2\rightarrow\infty$.
    The fit uses a shared timing offset and independent Rabi frequency, amplitude, and baseline for each microwave power.
    (b) Rabi frequencies $f_R=\Omega_R/2\pi$ extracted from the same simultaneous ensemble fit as a function of $\sqrt{P}$.
    The black line is a linear fit, $\Omega_R/2\pi=k\sqrt{P}$, to the three lower-power points, yielding $k=3.04$\,MHz/W$^{1/2}$.
    $T=5.5$\,K, $B_{\mathrm{ext}}=319$\,mT, and $f_{\mathrm{MW}}=8.20$\,GHz.
    }
    \label{fig:rabi}
\end{figure}

Having established hyperfine-resolved access to the donor spin system, we next investigate coherent microwave control by driving Rabi oscillations on the $m_I = -9/2$ transition. Fig.~\ref{fig:rabi}(a) shows Rabi oscillations for MW powers ranging from 20-107\,W. We observe coherent oscillations at all powers with  a maximum Rabi frequency of $\Omega_R/2\pi = 36.2 \pm 0.7$\,MHz. This corresponds to a $\pi$-pulse time of $t_\pi = 13.8 \pm 0.3$\,ns and a microwave magnetic-field amplitude of $B_{\mathrm{MW}} = 1.32 \pm 0.03$\,mT. The shortest $\pi$-pulse time is therefore comparable to, but shorter than, the inhomogeneous dephasing time ($T_2^*\approx17$\,ns) enabling coherent spin rotations despite the rapid ensemble dephasing. 

For the first three powers, the extracted Rabi frequencies increase linearly with the square root of the applied microwave power as shown in Fig.~\ref{fig:rabi}(b). There is a small deviation at the highest power suggesting amplifier nonlinearity. The scaling confirms that the observed oscillations arise from resonant microwave control of the donor electron spin. 

The Rabi decay time is significantly longer than $T_2^*$.  Resonant driving partially suppresses inhomogeneous dephasing, giving a Rabi damping time $T_{\mathrm{Rabi}}$ between $T_2^*$ and $T_2$. In the limit that $\Omega_R \gg \sigma$, the damping rate approaches $\sigma^2/\Omega$~\cite{abragam1961pnm}, while ultimately remaining limited by $T_2$. Our experiments lie in the intermediate regime. To test whether the observed Rabi dynamics are consistent with the measured ensemble dephasing, we numerically model the oscillations using a Gaussian detuning distribution ($T_2^* = 17$\,ns, $\sigma = 13$\,MHz) consistent with the PODMR linewidth and obtained with subsequent Ramsey measurements, and an infinite $T_2$. The resulting fits are shown in Fig.~\ref{fig:rabi}a, and reproduce the decay envelopes across all four microwave powers, demonstrating that ensemble inhomogeneity accounts for much of the observed damping. 

To obtain a conservative lower bound on $T_2$, we repeat the fit with the inhomogeneous broadening removed, such that all remaining decay is attributed to a finite $T_2$. This procedure gives 
$T_{\mathrm{Rabi}}=200\pm20$\,ns (App.~\ref{app:rabi_model_comparison}). Because any residual inhomogeneous broadening would also contribute to the observed decay, this value provides a lower bound on $T_2$. The short measurement window, however, does not permit a meaningful upper bound.

The ability to drive coherent spin rotations on timescales shorter than $T_2^*$ establishes microwave control of implanted donor electron spins despite substantial ensemble broadening present in natural abundance ZnO. Having demonstrated coherent control, we next investigate the coherence properties of the donor spin ensemble using Ramsey interferometry and dynamical decoupling measurements.


\section{Spin Coherence}
\begin{figure}[ht]
    \centering
    \includegraphics[width=0.85\linewidth]{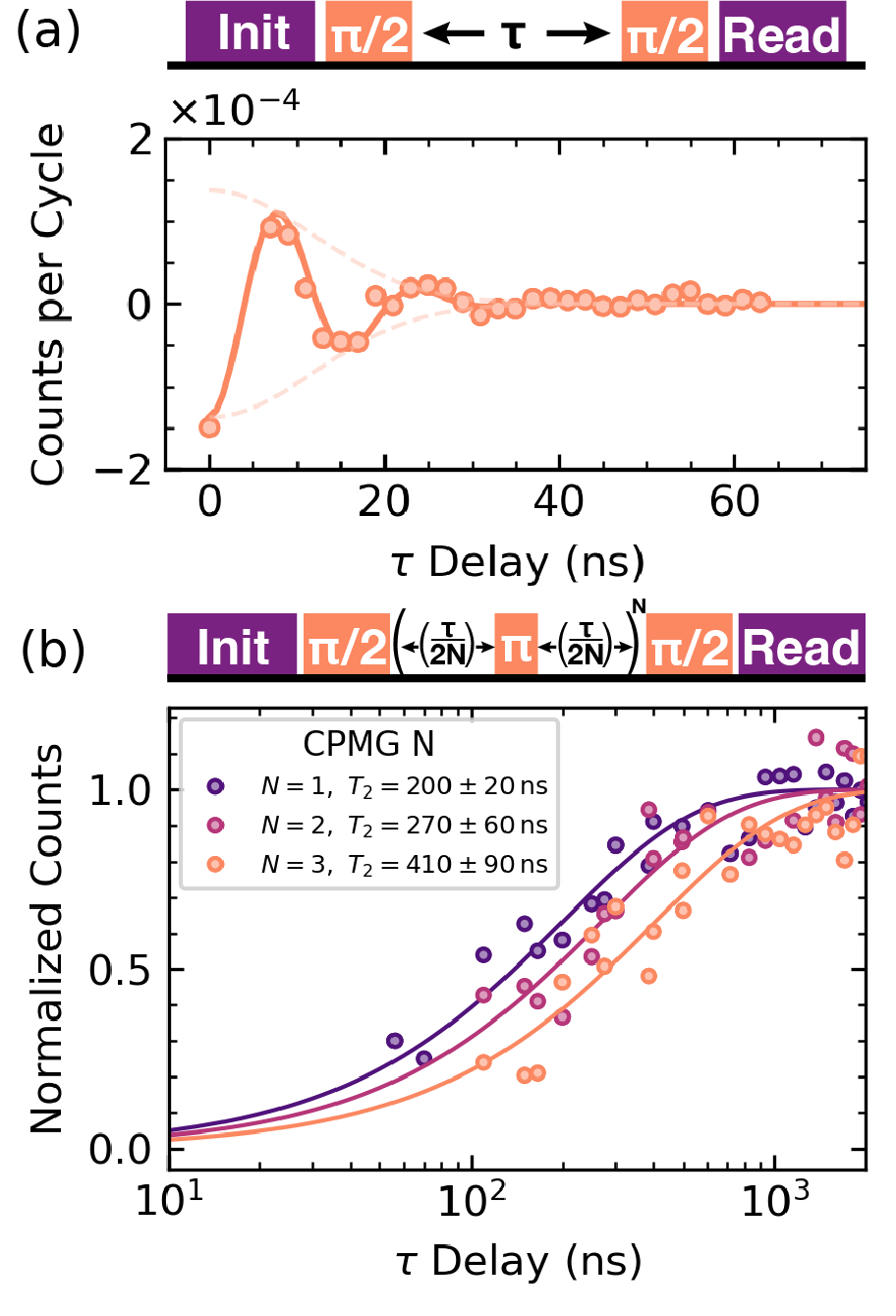}
    \caption{
    Spin coherence of implanted In donors in ZnO measured with \sigminus\ optical pumping.
    (a) Ramsey interferometry acquired with the microwave frequency detuned by
    $+60$\,MHz from the $m_I=-9/2$ transition ($B_{\mathrm{ext}}=311.1$\,mT). The data are fit to a damped
    oscillation,
    $S(\tau)=S_0 + A e^{(-\tau/T_2^*)^2}\cos(2\pi f_0 \tau+\phi)$,
    yielding $T_2^*=17\pm1$\,ns and $f_0=60\pm2$\,MHz, consistent with the
    applied microwave detuning.
    (b) Hahn-echo and CPMG measurements acquired on the $m_I=-9/2$ transition ($B_{\mathrm{ext}}=318.6$\,mT) for $N=1$, $2$,
    and $3$ refocusing pulses. The data are fit with a single exponential,
    $S_N(\tau)=S_{\infty,N}+A_N e^{-\tau/T_{2,N}}$. The extracted coherence times are listed
    in the legend. $\Omega_R/2\pi = 23$\,MHz, $T=5.5$\,K, and $f_{\mathrm{MW}}=8.20$\,GHz for all data.
    }
    \label{fig:coherence}
\end{figure}

Ramsey interferometry, shown in Fig.~\ref{fig:coherence}(a), yields an inhomogeneous dephasing time of $T_2^* = 17 \pm 1$\,ns. This value is consistent with the linewidth analysis of the PODMR spectra (App.~\ref{app:odmr_broadening}) and the Gaussian detuning distribution used to model the Rabi oscillations (App.~\ref{app:rabi_model_comparison}), supporting that all three measurements are described by the same inhomogeneous broadening. The measured $T_2^*$ is also comparable to the $17$\,ns value measured optically for Ga donors in ZnO~\cite{linpeng2018cps}. This agreement supports the interpretation that $T_2^*$ in natural ZnO is primarily limited by inhomogeneous hyperfine fields from the natural-abundance $^{67}$Zn nuclear-spin bath. Consequently, isotopic purification of the host lattice is expected to substantially extend the inhomogeneous dephasing time.

To extend the coherence beyond the inhomogeneous dephasing limit, we apply Carr--Purcell--Meiboom--Gill (CPMG) dynamical decoupling sequences, as shown in Fig.~\ref{fig:coherence}(b). For Hahn echo ($N=1$), we measure a coherence time of $T_2 = 200 \pm 20$\,ns. Increasing the number of refocusing pulses extends the coherence time, consistent with suppression of slowly varying magnetic field fluctuations. For $N>3$, however, substantial heating of the sample space is observed, preventing measurements at higher pulse numbers.

Although dynamical decoupling extends the coherence by more than an order of magnitude, 
the measured Hahn-echo coherence time is more than two orders of magnitude shorter than the $\sim 50~\mu$s value previously reported from all-optical measurements in ZnO~\cite{linpeng2018cps}. The discrepancy is unexpected given that the epilayer is grown by molecular-beam epitaxy and is expected to have lower concentrations of residual impurities than the bulk material used in Ref.~\cite{linpeng2018cps}. We therefore perform a series of control measurements to identify the origin of the shortened coherence.

We first test whether this reduction is caused by microwave-induced heating during the pulse sequences. Repeating the Hahn-echo measurement with the sample immersed in superfluid helium at $1.8$\,K, compared to 5.5\,K,  yields no measurable change in $T_2$ (App.~\ref{app:low_temp_echo}), indicating that a steady-state increase in the sample temperature is not the dominant limitation.

We next consider instantaneous diffusion, which arises when the microwave pulse flips spins dipole-coupled to the central spin, producing abrupt changes in the local magnetic field. For a randomly distributed spin ensemble, the instantaneous-diffusion-limited coherence time is 
\begin{equation}
\label{eq:t2id}
\frac{1}{T_{2,\mathrm{ID}}}
=
\frac{\mu_0 \pi (g\mu_B)^2 n_{\mathrm{eff}}}
{9\sqrt{3}\hbar}
\sin^2\left(\frac{\theta_2}{2}\right),
\end{equation}
where $n_{\mathrm{eff}}$ is the density of resonantly driven spins within the pulse bandwidth and $\theta_2$ is the refocusing-pulse angle~\cite{salikhov1981tes}. The pulse angle dependence has been used to experimentally identify and suppress instantaneous diffusion in donor-spin ensembles~\cite{tyryshkin2012esc}. 

Using the measured Hahn-echo coherence time, Eq.~\ref{eq:t2id} implies an effective density of resonantly driven spins of $n_{\mathrm{eff}}\sim 6\times 10^{18}~\mathrm{cm}^{-3}$, substantially larger than the peak implanted In concentration of $\sim 10^{16}~\mathrm{cm}^{-3}$ estimated from SRIM calculations (App.~\ref{app:samp_prep}). Reconciling the observed coherence time with instantaneous diffusion would therefore require a much larger density of additional resonantly driven paramagnetic centers. However, the measurements are performed on the $m_I = 9/2$ donor transition, corresponding to a resonance field well separated from the free-electron resonance and equivalent to an effective $g$ value of $\approx 1.85$. To our knowledge, no paramagnetic defect has been reported in ZnO at this resonance condition. These considerations already suggest that instantaneous diffusion is unlikely to account for the observed Hahn-echo decay. 

We test this conclusion directly by reducing the refocusing pulse angle from $\theta_2=\pi$ to $\theta_2=\pi/2$ (App.~\ref{app:refocus_angle_dep}). If instantaneous diffusion dominates the Hahn-echo decay, Eq.~\ref{eq:t2id} predicts a factor-of-two increase in $T_{2,\mathrm{ID}}$.  We measure statistically indistinguishable coherence times for the two cases, $T_2=230\pm40$~ns for $\theta_2=\pi$ and $T_2=240\pm30$~ns for $\theta_2=\pi/2$. Together with the observed extension under CPMG decoupling, these measurements indicate that instantaneous diffusion is not the dominant mechanism limiting the Hahn-echo coherence time. 

The control measurements above rule out microwave heating and instantaneous diffusion as the dominant limitations to the Hahn-echo coherence time. We therefore consider the remaining differences between the present study and the all-optical measurements of Ref.~\cite{linpeng2018cps}, including donor synthesis and species and magnetic-field regime.

The optical studies were performed on \textit{in situ} Ga donors unintentionally incorporated during growth in a Tokyo Denpa ZnO substrate at least 0.7\,\textmu m from the sample surface, due to an undoped capping MBE epilayer. The current study investigates In donors formed by implantation and annealing, with a peak donor depth of approximately 90\,nm below the surface. Since all three defects, In, Al and Ga, are shallow donors governed by effective mass physics, we do not expect a strong dependence of $T_2$ on species. We are able to perform Hahn echo measurements on Al donors in the substrate on the same sample. Further, we perform these measurements in regions that were masked from In implantation~(App.~\ref{app:al_hahn_echo}). We find the Al donor Hahn-echo times are also sub-microsecond, with $T_2\sim 100$--$400$~ns depending on position, and show no measurable improvement when the refocusing angle is reduced from $\pi$ to $\pi/2$ (App.~\ref{app:al_hahn_echo}). These measurements indicate that neither the In donor species nor the near-surface In implantation profile alone is responsible for the shortened coherence.

Taken together, the heating controls, instantaneous-diffusion measurements, and Al-donor comparison point toward magnetic noise from the broader local spin environment causing spectral diffusion as a dominant source of decoherence at the reduced magnetic field. A well-established mechanism in donor-spin systems is spectral diffusion arising from dipolar flip-flops within an electron-spin bath. When two bath spins exchange their spin states, the dipolar field experienced by the donor changes, shifting its resonance frequency. These stochastic frequency shifts are not fully refocused by a Hahn-echo sequence and lead to a finite coherence time~\cite{tyryshkin2012esc}. The observed increase in coherence time with the number of CPMG refocusing pulses is consistent with this picture, as additional refocusing pulses suppress dephasing arising from slowly varying bath fluctuations.

The magnetic-field regime may influence these bath dynamics. For an electron-spin bath at 5.5\,K, the thermal polarizations $P_{\mathrm{th}}$ are approximately 0.54 and 0.04 at 5\,T and 300\,mT, respectively. Since dipolar-mediated flip-flops require oppositely oriented spin pairs, the active pair fraction scales as $1-P_\mathrm{th}^2$, changing by only a factor of $\sim1.4$ between the two fields (App.~\ref{app:spin_bath_polarization}). Thermal polarization alone cannot explain the $\sim 250$-fold discrepancy between the present Hahn-echo coherence times and the $\sim50$\,\textmu s values reported previously. The higher magnetic field used in Ref.~\cite{linpeng2018cps}, however, may further suppress bath dynamics by increasing resonance-frequency mismatches between paramagnetic centers, thereby reducing flip-flop-mediated spectral diffusion. 

While the present measurements do not identify the microscopic origin of the fluctuating spins, they suggest that the local electronic spin environment plays a more important role than nuclear spins or instantaneous diffusion. Similarly short coherence times have been observed in Er-doped CeO$_2$~\cite{zhang2024osc}, while electronic-bath-driven spectral diffusion on the one microsecond timescale has been identified as the limiting mechanism in rare-earth spin ensembles~\cite{welinski2018esc}. Lower-temperature measurements could distinguish between thermally activated bath dynamics and more static sources of magnetic noise, while field-dependent studies could determine the extent to which spin-bath polarization and suppression of flip-flop processes contribute to the observed decoherence. 

\section{Conclusions}

We have demonstrated coherent microwave control of implanted $^{115}$In donor spins in ZnO, including hyperfine-resolved spin initialization and readout, Rabi oscillations with $\pi$-pulse times as short as $13.8$\,ns, and measurements of the donor-spin coherence using Ramsey, Hahn-echo, and CPMG pulse sequences. The measured inhomogeneous dephasing time, $T_2^*=17\pm1$\,ns, is consistently described by the PODMR linewidth, Ramsey measurements, and modeling of the Rabi decay, establishing a self-consistent picture of the donor-spin ensemble. These results establish implanted donors in ZnO as a platform for coherent microwave control and provide a foundation for future studies of donor-spin physics in this material system.

Unexpectedly, the Hahn-echo coherence time is limited to approximately $200$\,ns, more than two orders of magnitude shorter than previously reported all-optical measurements of donor spins in ZnO at high magnetic field. Through a series of control measurements, we rule out microwave heating and instantaneous diffusion as dominant explanations and find no evidence that donor species or near-surface implantation alone account for the shortened coherence. Instead, the results are consistent with spectral diffusion arising from the broader electronic spin environment. These observations suggest that, while nuclear spins govern the inhomogeneous dephasing time in natural ZnO, they do not set the ultimate coherence limit in the present experiments. More importantly, the microwave-control techniques demonstrated here provide direct access to the mechanisms limiting coherence and establish a platform for systematically studying the electronic spin environment in ZnO.

More broadly, these results highlight the importance of understanding and controlling electronic spin baths in emerging quantum defect host materials. Considerable effort has been devoted to identifying host materials with low nuclear-spin concentrations~\cite{kanai2022gss}, motivated by the success of isotopically purified silicon and diamond. Our measurements demonstrate that suppressing the nuclear-spin bath alone may be insufficient to realize long spin coherence times. Achieving robust donor-spin qubits in ZnO and other semiconductor platforms will require not only control of the nuclear-spin environment, but also identification and mitigation of residual paramagnetic defects and impurities that contribute to spectral diffusion. The combination of optical spin initialization and readout with fast microwave control demonstrated here, which is applicable to both implanted ensembles and single centers, provides a direct route to correlating coherence with magnetic field, temperature, isotopic composition, defect density, and materials processing. Such studies will be essential for identifying the dominant sources of decoherence and guiding the development of improved quantum defect materials.

\section*{Acknowledgments}
The authors thank Stefan Stoll for helpful discussions on forbidden transitions in ODMR, Matt Reynolds for discussion on MW resonator design, and David Pederson at the Center for Experimental Nuclear and Astrophysics for his assistance with resonator fabrication.
The work was primarily supported by the National Science Foundation under Grant No. 2212017. Donor formation and modeling was supported by the AFOSR CFIRE program under grant FA9550-23-1-0418. Additional support from JSPS Grants-in-Aid for Scientific Research (S) Grant No. JP22H04958 (M.K.).

\appendix
\renewcommand{\thefigure}{A\arabic{figure}}
\setcounter{figure}{0}
\section{Spin Initialization Fidelity}
\label{app:spin_init}

To account for laser scatter in the optical-pumping measurements, we compare resonant and off-resonant pulse sequences under otherwise identical conditions. Figure~\ref{fig:opticalAccess}(a) shows optical-pumping transients acquired with the scramble pulse applied and the resonant pump tuned either on resonance with the donor-bound-exciton transition or off resonance near the donor peak. When the pump is detuned from the optical transition, optical pumping is suppressed, and the measured signal is dominated by laser scatter and nonresonant background. From this off-resonant trace, we determine a background level of approximately $2.3\times10^{4}$ counts. This value is subtracted from the resonant optical-pumping signal before normalizing the data in Fig.~\ref{fig:opticalAccess}(b).

\begin{figure}[ht]
    \centering
    \includegraphics[width=0.95\linewidth]{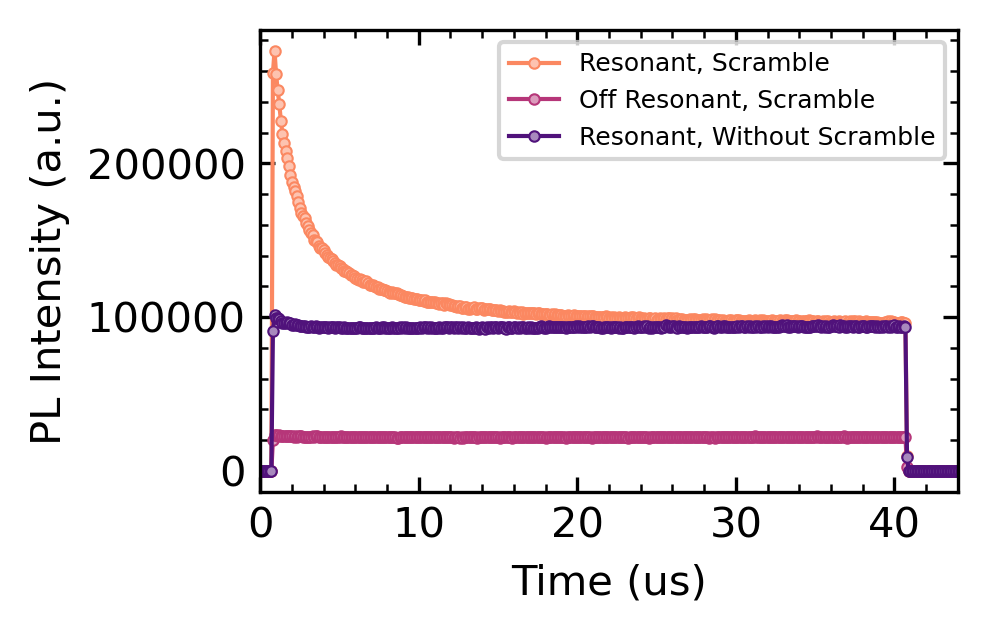}
    \caption{
    Laser-scatter background characterization for optical-pumping measurements. Time-resolved PL transients acquired at $T=5.5$\,K and $B=300$\,mT are shown for resonant optical pumping with the scramble pulse applied, off-resonant pumping with the scramble pulse applied, and resonant pumping with the scramble pulse blocked. The off-resonant trace provides an estimate of the laser-scatter background, for which optical pumping is suppressed and the detected signal is dominated by nonresonant background.
    }
    \label{fig:appendix_spin_init_background}
\end{figure}

As a consistency check, we also repeat the resonant optical-pumping measurement with the scrambling pulse blocked. In this case, the spin population is not randomized before the pump pulse. Because the delay between successive pump pulses is much shorter than $T_1$, little optical-pumping contrast is observed. The trace approaches the same long-time equilibrium signal as the resonant measurement with the scrambling pulse as expected.

\section{PODMR Envelope}
\label{app:odmr_bowl-shape}

We now consider which terms in the spin Hamiltonian could give rise to the angle-dependent bowl-shaped envelope. In frequency units, the static Hamiltonian is written as
\begin{equation}
\frac{\mathcal{H}_0}{h} = \frac{\mu_B}{h}\,\mathbf{B}_0\cdot \mathbf{g}\cdot \mathbf{S} - \gamma_n\,\mathbf{B}_0\cdot \mathbf{I} + \mathbf{S}\cdot \frac{\mathbf{A}}{h}\cdot \mathbf{I} + \frac{P_\parallel}{h}\left[I_{z_c}^2-\frac{I(I+1)}{3}\right],
\label{eq:in_spin_hamiltonian}
\end{equation}
in which $\gamma_n = 0.9329~\mathrm{kHz/G}$. Both $\mathbf{g}$ and $\mathbf{A}$ are reported to be highly isotropic ($g_\parallel = 1.9574$, $g_\perp = 1.9562$, $\gamma_n = 0.9329~\mathrm{kHz/G}$; $A_\parallel/h = 100.28~\mathrm{MHz}$, and $A_\perp/h = 100.14~\mathrm{MHz}$)~\cite{block1982odm} which is consistent with our measurements. $P_\parallel$ is reported to be 1.27\,MHz~\cite{block1982odm}. Moreover, if $g$ and $A$ were significantly anisotropic, tilting $\mathbf{B}_0$ away from the $c$ axis would change the effective hyperfine splitting, so the measured peak spacing would deviate from the isotropic prediction and vary with the misalignment angle. The spacing we observe is instead well described by nearly isotropic $g$ and $A$, with no such angular variation. 

The only remaining intrinsic term in the spin Hamiltonian that could contribute to the bowl-shape feature is the nuclear quadrupole interaction. In our experimental regime, the main effect of the quadrupole interaction is the appearance of forbidden transitions with field angle. At angles at which the forbidden transitions become significant, the transitions spacing also becomes non-uniform since the quadrupole shift scales as $m_I^2$. Introducing a larger-than-reported quadrupole interaction also cannot reproduce the spectrum as the forbidden transitions become appreciable. Neither non-uniform spacing nor strong forbidden transitions are observed experimentally.  
The quadrupole interaction can therefore be excluded as the origin of the envelope. We thus conclude that the intrinsic anisotropy of the spin Hamiltonian ($g$, $A$, and the quadrupole interaction) cannot account for the bowl-shaped envelope or its angular dependence. A complete account of the envelope and its dependence on the B-field alignment is left to future work.

\section{Lower bound on $T_2$ from Rabi Measurements}
\label{app:rabi_model_comparison}

In the main text, the Rabi oscillations are modeled by averaging the driven two-level response over a fixed Gaussian distribution of static microwave detunings, with width $\sigma_\Delta/2\pi=13$\,MHz. This model accounts for the inhomogeneous broadening inferred from the PODMR linewidth and Ramsey measurements, and reproduces much of the observed decay without requiring an additional homogeneous decay channel.

\begin{figure}[ht]
    \centering
    \includegraphics[width=0.85\linewidth]{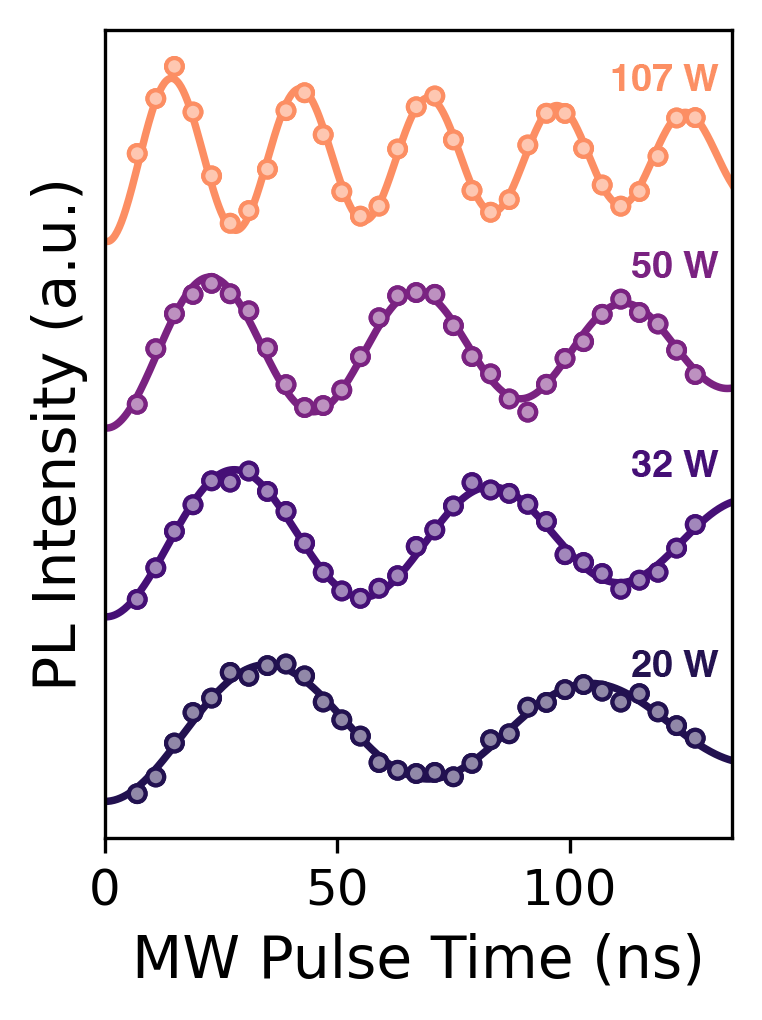}
    \caption{
    Homogeneous-decay-only fit to the Rabi data.
    The same Rabi oscillations shown in Fig.~\ref{fig:rabi} are fit with the inhomogeneous detuning distribution removed, $\sigma_\Delta=0$, so that the observed damping is attributed entirely to a finite homogeneous coherence time.
    The pulse sequence consists of resonant optical initialization, a variable-duration microwave pulse, and resonant optical readout.
    The traces are vertically offset for clarity.
    The fit is performed simultaneously across all microwave powers with a shared $T_2$ and independent Rabi frequency, amplitude, and offset for each trace, yielding $T_2 = 200\pm20$\,ns.
    }
    \label{fig:app_rabi_sigma0}
\end{figure}

To place a conservative lower bound on the homogeneous coherence time, we repeat the simultaneous fit after removing the detuning distribution, setting $\sigma_\Delta=0$. In this comparison model, all observed damping of the Rabi oscillations is instead attributed to a finite homogeneous coherence time $T_2$. The same Rabi traces are fit simultaneously across microwave powers with a shared $T_2$, while allowing the Rabi frequency, amplitude, and offset to vary independently for each power. The resulting fit is shown in Fig.~\ref{fig:app_rabi_sigma0} and gives $T_2 = 200\pm20~\mathrm{ns}$.

This value should not be interpreted as a direct measurement of the intrinsic Hahn-echo coherence time. Rather, it is a conservative lower bound on the homogeneous coherence time relevant to the driven Rabi dynamics. Any residual inhomogeneous broadening, pulse imperfections, or power-dependent technical damping would also contribute to the observed decay; neglecting these effects forces the fit to attribute the full decay envelope to finite $T_2$, thereby underestimating the homogeneous coherence time.

\section{PODMR Linewidth}
\label{app:odmr_broadening}

The PODMR spectra in Fig.~\ref{fig:odmr}(a,b) are acquired using a square microwave $\pi$ pulse. 
Even in the absence of additional inhomogeneous broadening, a finite pulse duration produces a finite excitation bandwidth. 
For an isolated two-level transition driven with on-resonance Rabi frequency $f_R=\Omega_R/2\pi$ and detuning $\delta$ (in linear-frequency units), the transition probability after a pulse of duration $t_p$ is

\begin{equation}
P(\delta,t_p)=
\frac{f_R^2}{f_R^2+\delta^2}
\sin^2\!\Big(\pi t_p\sqrt{f_R^2+\delta^2}\Big).
\label{eq:rabi_lineshape_app}
\end{equation}

For a resonant $\pi$ pulse, $t_p=t_\pi=1/(2f_R)=\pi/\Omega_R$, giving

\begin{equation}
P_\pi(\delta)=
\frac{1}{1+(\delta/f_R)^2}
\sin^2\!\left[
\frac{\pi}{2}\sqrt{1+(\delta/f_R)^2}
\right].
\label{eq:rabi_pi_lineshape_app}
\end{equation}

The corresponding full width at half maximum is

\begin{equation}
\Gamma_R^{(\nu)} \approx \frac{0.799}{t_\pi}
=1.60\,f_R
=1.60\,\frac{\Omega_R}{2\pi},
\label{eq:rabi_fwhm_app}
\end{equation}

where $\Gamma_R^{(\nu)}$ is expressed in linear-frequency units. 
Using $\Omega_R/2\pi=23$\,MHz gives $t_\pi=21.7$\,ns and

\begin{equation}
\Gamma_R^{(\nu)} \approx 36.8~\mathrm{MHz}.
\end{equation}

The measured inhomogeneous dephasing time is dominated by the quasi-static nuclear spin bath, for which the Ramsey envelope is well described by a Gaussian decay,

\begin{equation}
C(\tau)=\exp\left[-\left(\frac{\tau}{T_2^*}\right)^2\right].
\end{equation}

This corresponds to a Gaussian distribution of spin-transition detunings,

\begin{equation}
G_{T_2^*}(\delta)=
\frac{1}{\sigma_\delta\sqrt{2\pi}}
\exp\left[-\frac{\delta^2}{2\sigma_\delta^2}\right],
\qquad
\sigma_\delta=\frac{1}{\sqrt{2}\pi T_2^*}.
\label{eq:t2star_gaussian_kernel_app}
\end{equation}

The associated full width at half maximum is

\begin{equation}
\Gamma_{T_2^*}^{(\nu)}
=
2\sqrt{2\ln 2}\,\sigma_\delta
=
\frac{2\sqrt{\ln 2}}{\pi T_2^*}.
\label{eq:t2star_gaussian_fwhm_app}
\end{equation}

Using $T_2^*=17$\,ns gives

\begin{equation}
\Gamma_{T_2^*}^{(\nu)} \approx 31.2~\mathrm{MHz}.
\end{equation}

A simple estimate for the expected PODMR linewidth is therefore obtained by convolving the finite-bandwidth $\pi$-pulse response with the Gaussian detuning distribution associated with $T_2^*$,

\begin{equation}
S(\delta)=
\int_{-\infty}^{\infty}
P_\pi(\delta-\delta')\,G_{T_2^*}(\delta')\,d\delta'.
\label{eq:odmr_convolution_app}
\end{equation}

For the present parameters, the finite $\pi$-pulse bandwidth contributes $\Gamma_R^{(\nu)}\approx 36.8$\,MHz and the measured inhomogeneous dephasing corresponds to a Gaussian linewidth $\Gamma_{T_2^*}^{(\nu)}\approx 31.2$\,MHz. Numerical evaluation of Eq.~\ref{eq:odmr_convolution_app} gives an expected convolved linewidth of

\begin{equation}
\Gamma_{\mathrm{conv}}^{(\nu)} \approx 46.7~\mathrm{MHz}.
\end{equation}

This value is consistent with the typical measured PODMR peak linewidth of $\sim 46$\,MHz.

\section{Hahn echo temperature dependence}

\label{app:low_temp_echo}

\begin{figure}[ht]
    \centering
    \includegraphics[width=0.95\linewidth]{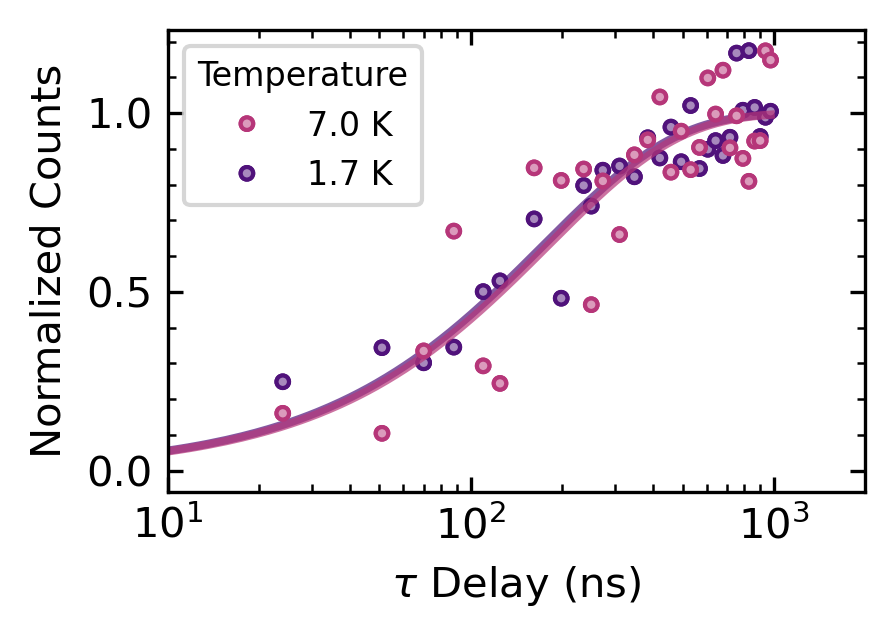}
    \caption{
    Low-temperature Hahn-echo comparison. 
    Hahn-echo measurements acquired on the $m_I=-9/2$ transition ($B_{\mathrm{ext}}=318.6$\,mT) and measured with \sigminus\ optical pumping at the same spatial location for a nominal sample-space temperature of $7.0$\,K and with the sample immersed in liquid helium at $1.7$\,K. $\Omega_R/2\pi = 19$\,MHz and $f_{\mathrm{MW}}=8.20$\,GHz for both measurements. The data are fit with a single exponential,
    $S_N(\tau)=S_{\infty,N}+A_N e^{-\tau/T_{2,N}}$, yielding a $T_2 = 200 \pm 40$\,ns and $T_2 = 210 \pm 40$\,ns for $7.0$\,K and $1.7$\,K, respectively.
    }
    \label{fig:hahn_echo_low_temp}
\end{figure}

To test whether the short Hahn-echo coherence time is caused by microwave-induced heating, we repeat the Hahn-echo measurement at the same spatial location at two temperatures, as shown in Fig.~\ref{fig:hahn_echo_low_temp}. The first measurement is performed at a sample-space temperature of $7.0$\,K, while the second is performed with the sample immersed in liquid helium at $1.7$\,K. In the immersed configuration, strong resonator-induced heating would be expected to produce visible boiling or turbulent helium flow in the sample space during the pulse sequence; no such behavior is observed during the measurement. 

The Hahn-echo decays measured under the two conditions show no resolvable difference in the extracted coherence time. This lack of temperature dependence indicates that the measured $T_2$ is not limited by a simple increase in the equilibrium sample temperature caused by microwave dissipation. 

\section{Hahn echo pulse area dependence}
\label{app:refocus_angle_dep}

To test whether the measured Hahn-echo decay is limited primarily by instantaneous diffusion, we compare echo measurements using different refocusing-pulse angles. Instantaneous diffusion arises from pulse-induced changes to the dipolar field produced by resonantly driven spins, and its contribution to the decay rate is expected to scale as $\sin^2(\theta/2)$, where $\theta$ is the rotation angle of the refocusing pulse. Reducing the refocusing pulse from $\pi$ to $\pi/2$ should therefore suppress the instantaneous-diffusion contribution and increase the measured coherence time if instantaneous diffusion is the dominant decay mechanism.

\begin{figure}[ht]
    \centering
    \includegraphics[width=0.95\linewidth]{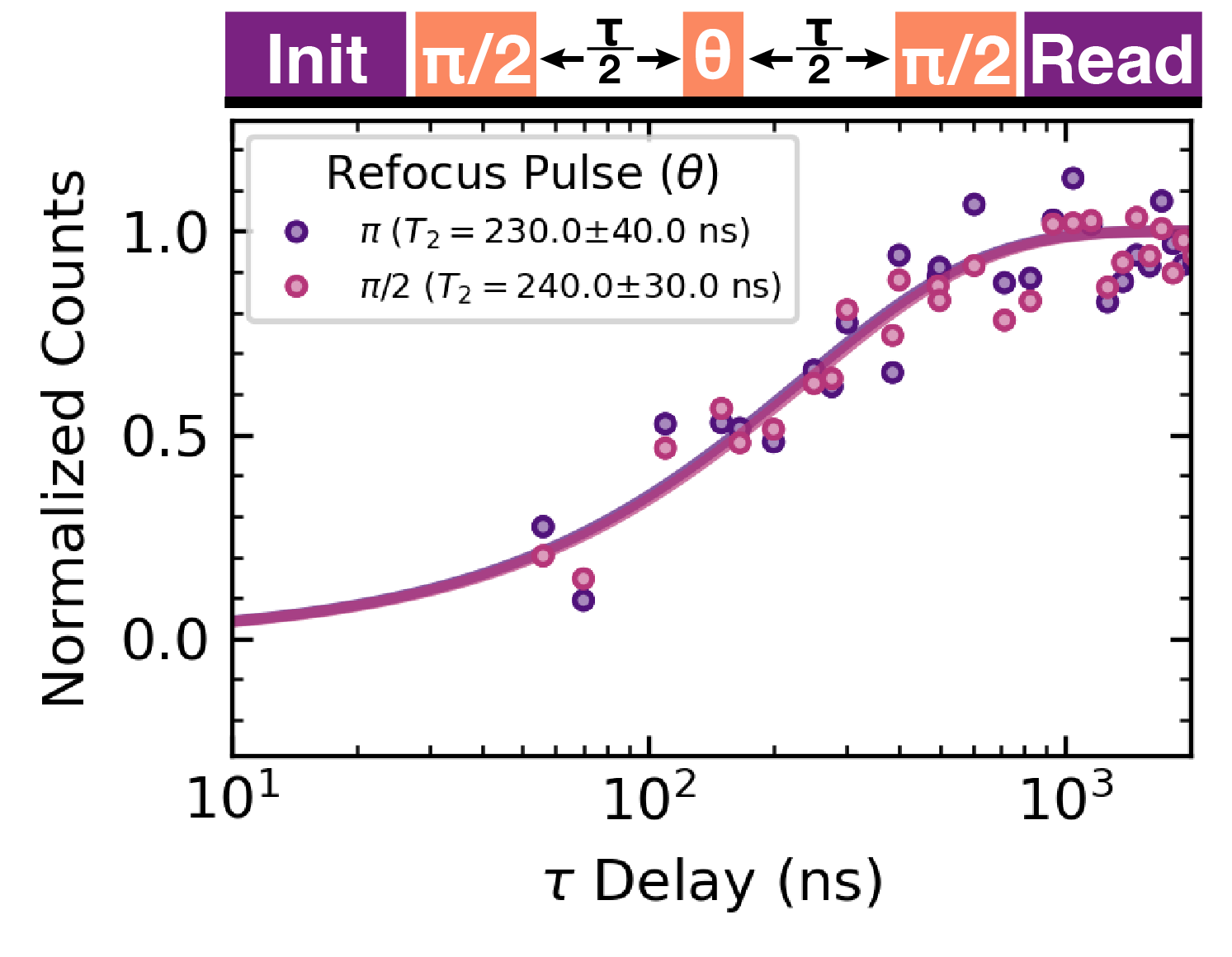}
    \caption{
    Refocusing-pulse-angle dependence of the Hahn-echo decay.
    Hahn-echo measurements at different refocusing pulse angles acquired on the $m_I=-9/2$ transition ($B_{\mathrm{ext}}=318.6$\,mT) and measured with \sigminus\ optical pumping. The pulse sequence consists of optical initialization, a microwave $\pi/2$ pulse, free evolution for $\tau/2$, a refocusing pulse with rotation angle $\theta$, a second $\tau/2$ evolution period, and a final microwave $\pi/2$ pulse before optical readout. Echo decays are shown for $\theta=\pi$ and $\theta=\pi/2$. The extracted coherence times are $T_2=230\pm40$\,ns and $T_2=240\pm30$\,ns, respectively, showing no measurable enhancement for the reduced refocusing-pulse angle. $\Omega_R/2\pi = 19$\,MHz and $T=5$\,K.
    }
    \label{fig:small_angle_echo}
\end{figure}

The pulse sequence used for this comparison is shown in Fig.~\ref{fig:small_angle_echo}. After optical initialization, a microwave $\pi/2$ pulse prepares a spin coherence, followed by a delay $\tau/2$, a refocusing pulse with angle $\theta$, a second delay $\tau/2$, and a final microwave $\pi/2$ pulse before optical readout. We compare measurements with $\theta=\pi$ and $\theta=\pi/2$ under otherwise identical conditions. The extracted coherence times are $T_2 = 230 \pm 40$\,ns for $\theta=\pi$ and $T_2 = 240 \pm 30$\,ns for $\theta=\pi/2$, showing no measurable enhancement for the reduced refocusing-pulse angle.

This lack of improvement indicates that the Hahn-echo decay is not dominated by instantaneous diffusion from the resonantly driven donor ensemble or resonantly driven environmental spins. This is supported by the extension in $T_2$ observed in the CPMG sequences, Fig.~\ref{fig:coherence}(b).

\section{Hahn echo on $in situ$-doped aluminum donors}
\label{app:al_hahn_echo}
To test whether the short Hahn-echo coherence time measured for implanted In donors is due to a property of the implanted donors, such as implantation damage or proximity to the surface, we repeat Hahn-echo measurements on residual Al donors in a region of the sample that was not implanted with In. The Al donor signal is expected to originate predominantly from donors below the high-purity epilayer, providing a comparison to an optically addressable donor ensemble that is not directly associated with the near-surface In implantation profile.

\begin{figure}[ht]
    \centering
    \includegraphics[width=0.95\linewidth]{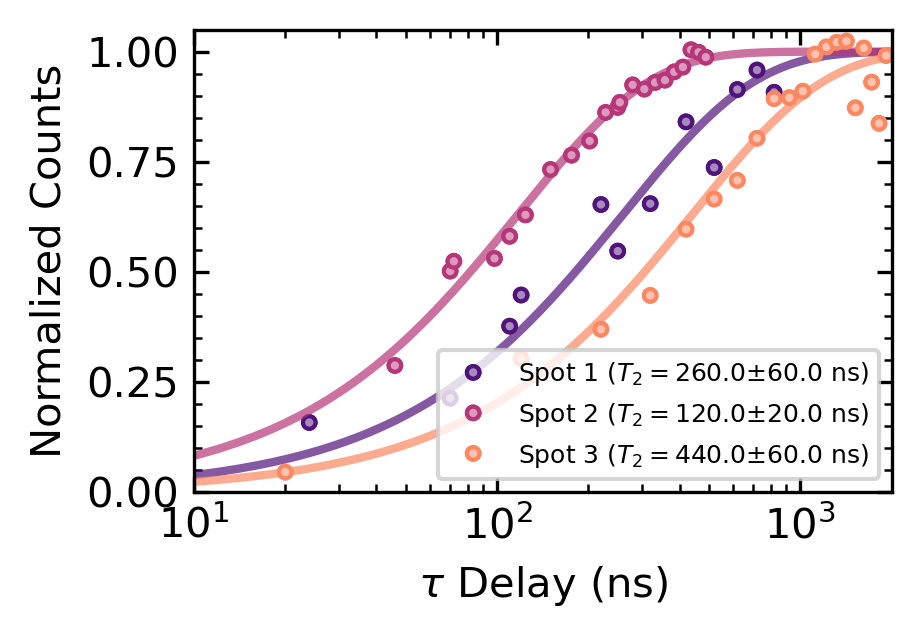}
    \caption{
    Hahn-echo measurements of residual Al donors acquired at three spatially separated locations in a region of the ZnO sample that was not implanted with In.
    Solid curves are fits to the Hahn-echo envelope, yielding $T_2=260\pm60$\,ns, $120\pm20$\,ns, and $440\pm60$\,ns for Spots 1--3, respectively.
    $\Omega_R/2\pi = 19$\,MHz and $T=1.8$\,K.
    }
    \label{fig:al_hahn_echo}
\end{figure}

Figure~\ref{fig:al_hahn_echo} shows Hahn-echo measurements acquired at three spatially separated locations. The extracted coherence times range from $T_2=120\pm20$\,ns to $440\pm60$\,ns. Thus, sub-microsecond Hahn-echo coherence is also observed in a non-implanted region of the sample. This comparison indicates that the short $T_2$ measured for implanted In donors is unlikely to arise solely from In implantation damage or from the shallow depth of the implanted donor ensemble.

The Al donor measurements also exhibit substantial spot-to-spot variation. This spatial dependence suggests that the local magnetic-noise environment varies across the sample, potentially due to inhomogeneous concentrations of residual donors or other paramagnetic defects. 

\begin{figure}[ht]
    \centering
    \includegraphics[width=0.95\linewidth]{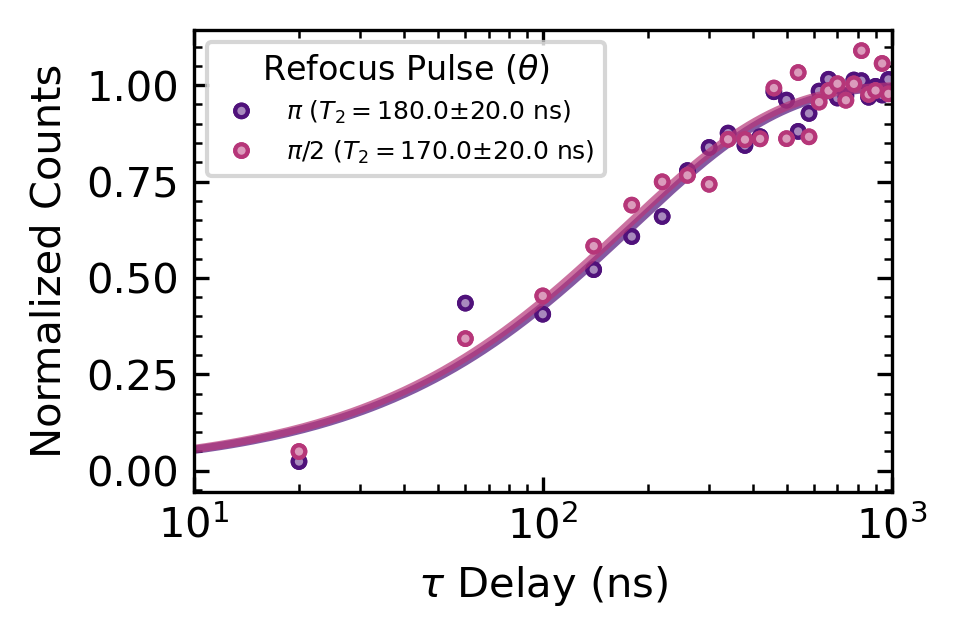}
    \caption{
    Refocusing-angle dependence of the Hahn-echo decay measured on residual Al donors in an unimplanted region of the ZnO sample.
    The echo decay is measured using a standard $\pi$ refocusing pulse and a reduced-angle $\pi/2$ refocusing pulse.
    Solid curves are fits to the echo envelope, yielding $T_2=180\pm20$\,ns for $\theta_2=\pi$ and $T_2=170\pm20$\,ns for $\theta_2=\pi/2$.
    Because the instantaneous-diffusion contribution to the decay rate scales as $\sin^2(\theta_2/2)$, reducing the refocusing angle from $\pi$ to $\pi/2$ should reduce this contribution by a factor of two. $\Omega_R/2\pi = 19$\,MHz and $T=1.8$\,K.
    }
    \label{fig:al_hahn_echo_angle_dep}
\end{figure}

As an additional control, we compare Al donor Hahn-echo measurements acquired with $\pi$ and $\pi/2$ refocusing pulses, shown in Fig.~\ref{fig:al_hahn_echo_angle_dep}. The extracted coherence times are $T_2=180\pm20$\,ns for $\theta_2=\pi$ and $T_2=170\pm20$\,ns for $\theta_2=\pi/2$. As discussed in App.~\ref{app:refocus_angle_dep}, if instantaneous diffusion dominated the Al donor Hahn-echo decay, reducing the refocusing pulse from $\pi$ to $\pi/2$ would reduce the instantaneous-diffusion contribution to the decay rate by a factor of two. Experimentally, we observe no such extension.

Together, the observation of short Al donor coherence in a non-implanted region, the spatial variation of the Al donor $T_2$, and the lack of refocusing-angle dependence suggest that the reduced coherence in the present sample reflects a broader magnetic-noise environment rather than a mechanism unique to the implanted In donor ensemble.

\section{Electron-spin-bath polarization estimate}
\label{app:spin_bath_polarization}

Here we estimate how the magnetic-field regime changes the thermal polarization of an electron-spin bath. For an electron spin in a magnetic field, the thermal polarization is
\begin{equation}
P(B,T)=\tanh\left(\frac{g\mu_B B}{2 k_B T}\right),
\end{equation}
where $P=1$ corresponds to complete polarization into the lower-energy spin state. For a bath of independent spin-$1/2$ moments, the populations of the two spin states are $
p_{\uparrow,\downarrow}=(1\pm P)/2.$
Dipolar-mediated flip-flops require oppositely oriented spin pairs. The fraction of anti-aligned pairs is therefore $
p_{\uparrow}p_{\downarrow}+p_{\downarrow}p_{\uparrow}
= (1-P^2)/2.
$

Relative to an unpolarized bath, the number of available flip-flop pairs is reduced by the factor
\begin{equation}
F_{\mathrm{ff}}(B,T)=1-P(B,T)^2.
\end{equation}
At $T=5.5~\mathrm{K}$, this gives
\begin{equation}
P(5~\mathrm{T}) \simeq 0.54,\qquad
P(0.3~\mathrm{T}) \simeq 0.04,
\end{equation}
and therefore
\begin{equation}
F_{\mathrm{ff}}(5~\mathrm{T})\simeq 0.71,\qquad
F_{\mathrm{ff}}(0.3~\mathrm{T})\simeq 1.00.
\end{equation}
The number of available flip-flop pairs is therefore larger at $300~\mathrm{mT}$ than at $5~\mathrm{T}$ by approximately
\begin{equation}
\frac{F_{\mathrm{ff}}(0.3~\mathrm{T})}{F_{\mathrm{ff}}(5~\mathrm{T})}
\simeq 1.4.
\end{equation}

This estimate shows that the lower magnetic field used in the present measurements places a possible electron-spin bath closer to the unpolarized limit. This effect is in the correct direction to increase flip-flop-driven spectral diffusion, but the magnitude of the polarization change alone is modest. Therefore, the field regime may contribute to the shorter Hahn-echo coherence time measured here, but it is unlikely to provide a complete microscopic explanation from a pure polarization standpoint. Further measurements of $T_2$ as a function of magnetic field, temperature, and donor or defect density will be needed to identify the dominant magnetic-noise source.

\section{Experimental Details}
\label{app:experiment}

All measurements were performed in a helium immersion cryostat equipped with a superconducting magnet capable of fields up to $7$\,T. Unless otherwise specified, experiments were carried out at a base temperature of $5$\,K; by pumping on the helium bath, the sample space can be cooled to approximately $1.8$\,K.

\begin{figure*}[ht]
    \centering
    \includegraphics[width=0.8\linewidth]{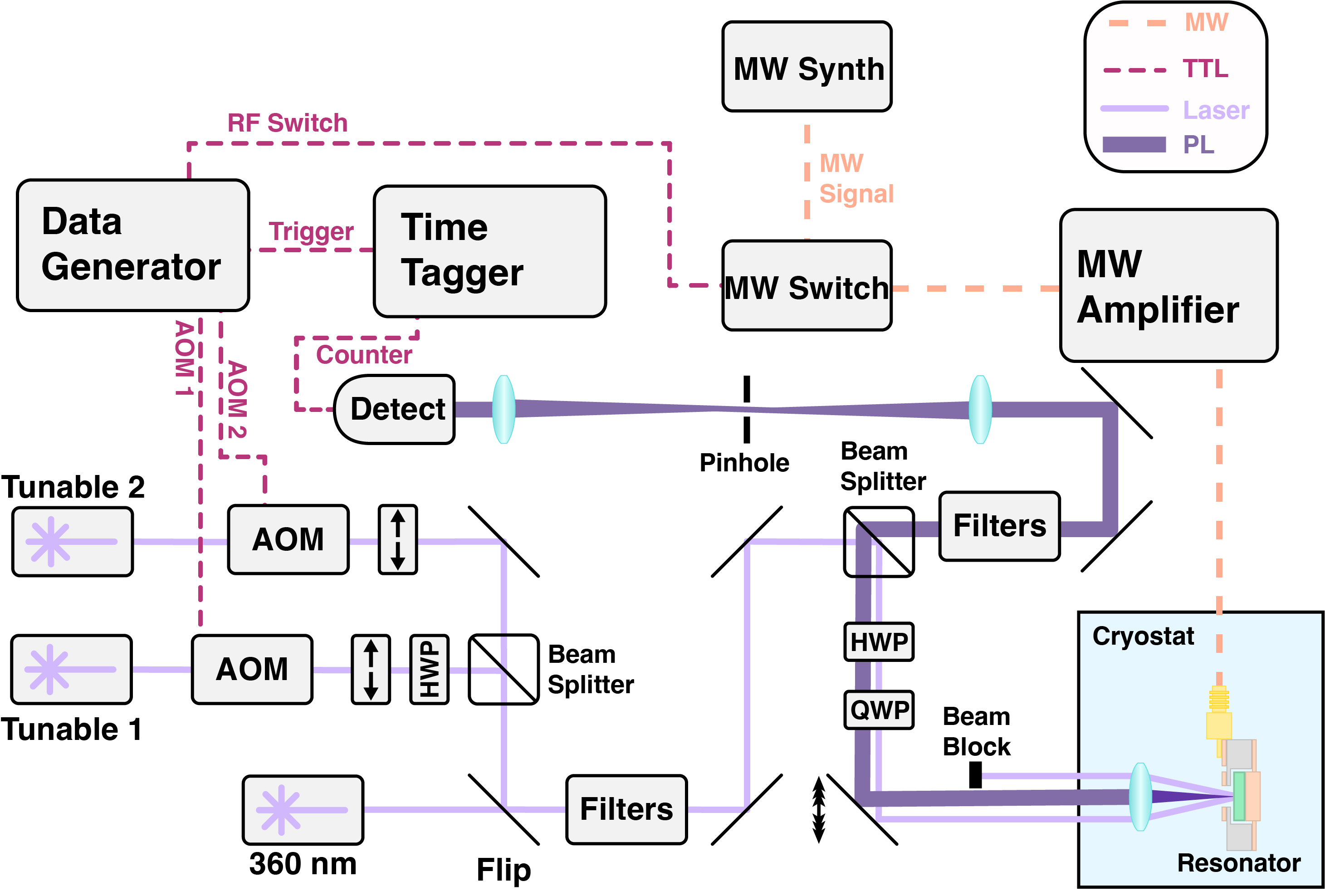}
    \caption{Diagram of Experimental Apparatus}
    \label{fig:setup}
\end{figure*}

Optical excitation and collection were performed using the home-built confocal microscope shown schematically in Fig.~\ref{fig:setup}. Above-band photoluminescence (PL) measurements were performed using a $360$\,nm diode laser (CNI MSL-F-360). For resonant measurements, we used two tunable ultraviolet lasers: a Spectra-Physics Matisse-S and a Toptica DL Pro HP-029344. Each tunable laser was independently gated using an acousto-optic modulator (AOM; Gooch and Housego 15210), with TTL control signals generated by a digital pattern generator (Wavepond DPG11). The two resonant excitation paths were then combined on a polarizing beam splitter with orthogonal linear polarizations.

To enable resonant excitation and collection on the same donor-bound-exciton transition, we used a side-excitation geometry. In this configuration, the excitation beam is displaced to the outer edge of the objective aperture using a mirror mounted on a translation stage. The beam is then focused onto the sample, while the reflected laser light is recollimated by the objective on the opposite side of the aperture. In contrast, the donor PL fills the objective more uniformly. A beam block placed in the reflected-laser path suppresses the reflected laser background while allowing the majority of the PL to be collected.

The collected PL was detected using a single-photon counting module (SPCM; PicoQuant PMA Hybrid 06), whose output was recorded with a Swabian Time Tagger Ultra. For triggered detection measurements, photon arrival times were recorded relative to the experimental pulse sequence, allowing time-resolved PL histograms to be constructed for optical pumping, pulsed ODMR, Ramsey, and echo measurements.

Microwave signals were generated using a microwave/RF signal generator (Windfreak SynthHD Pro V2). The continuous microwave output was gated using a fast microwave switch (Quantic PMI P1T-DC40G-65-T-292FF-1NS-OPT18G) driven by the same digital pattern generator used for the optical pulse sequence. The pulsed microwave signal was then amplified using a high-power traveling-wave-tube amplifier (TWT; Instruments for Industry T186-50). The amplified microwave pulses were delivered to the resonator mounted inside the cryostat with an SMA cable.

\section{Sample Preparation}
\label{app:samp_prep}

The ZnO epilayer surface was patterned during implantation using a square TEM grid as a shadow mask. The grid had a pitch of $51\,\text{\textmu m}$, with $28\,\text{\textmu m}$-wide openings separated by $23\,\text{\textmu m}$-wide bars. Indium ions were implanted through the mask at a dose of $10^{11}\,\mathrm{ions/cm^2}$ and an energy of $380$\,keV. Implantation was performed by Cutting Edge Ions.

\begin{figure}[ht]
    \centering
    \includegraphics[width=0.9\linewidth]{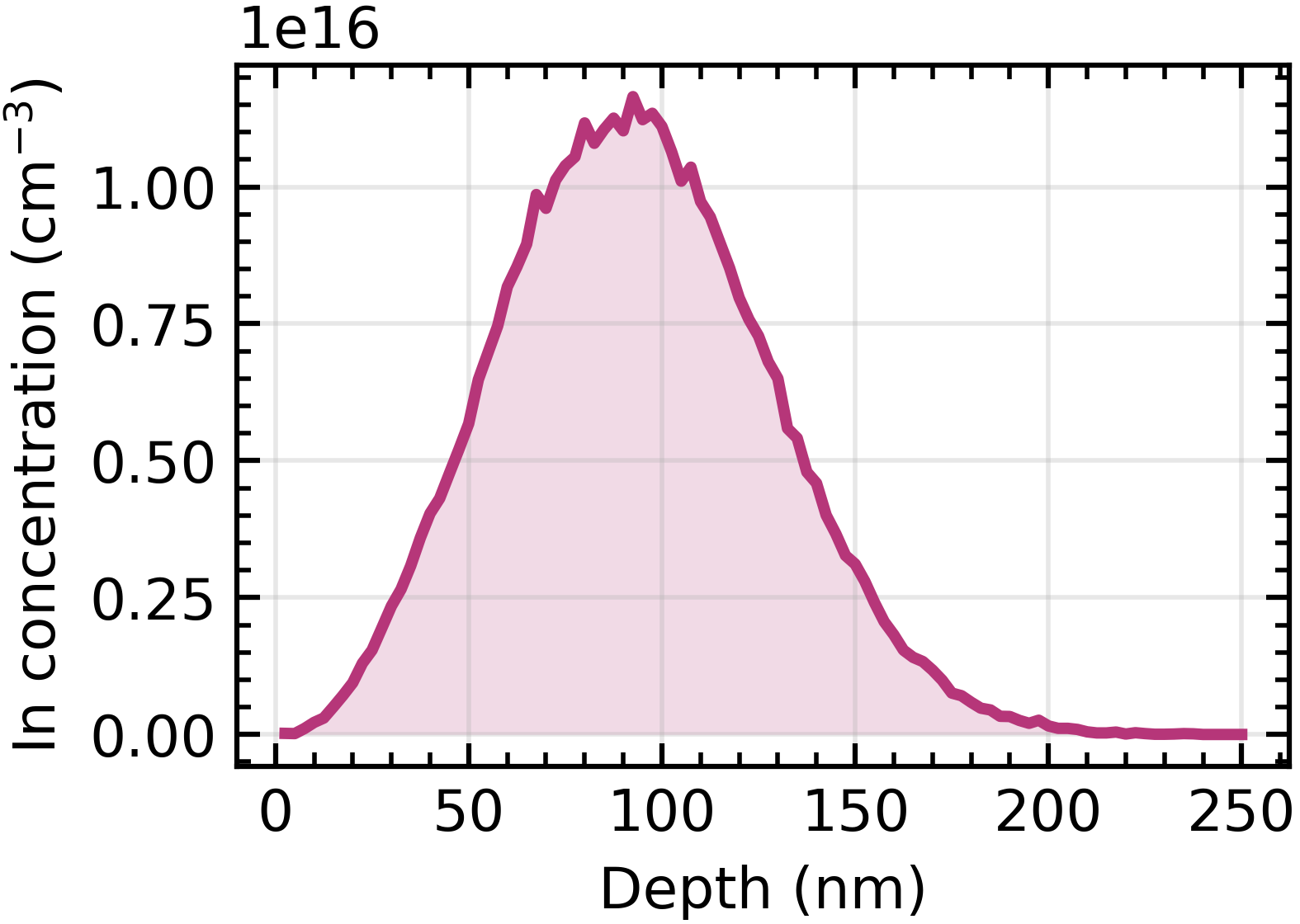}
    \caption{
    Simulated depth distribution of implanted In ions in ZnO calculated using SRIM for the implantation conditions used in this work. 
    The implanted profile is concentrated within the near-surface region, with a peak depth of approximately $90$-$100$~nm and a maximum In concentration of $\sim 1\times 10^{16}~\mathrm{cm^{-3}}$. 
    The distribution falls rapidly beyond $\sim 150$~nm, indicating that the optically addressed donor ensemble is localized within the first few hundred nanometers of the ZnO surface.
    }
    \label{fig:srim}
\end{figure}

This process produced an array of nominally $28\,\text{\textmu m} \times 28\,\text{\textmu m}$ implanted regions separated by unimplanted areas. SRIM simulations indicate that the implanted In distribution is centered at a depth of approximately $100$\,nm below the surface, Fig.~\ref{fig:srim}. Following implantation, the sample was annealed at $700\,^\circ\mathrm{C}$ for $1$\,h under \Otwo\ flow to reduce implantation-induced damage.

\section{Optical Characterization}
\label{app:optical_characterization}

\begin{figure}[ht]
    \centering
    \includegraphics[width=0.9\linewidth]{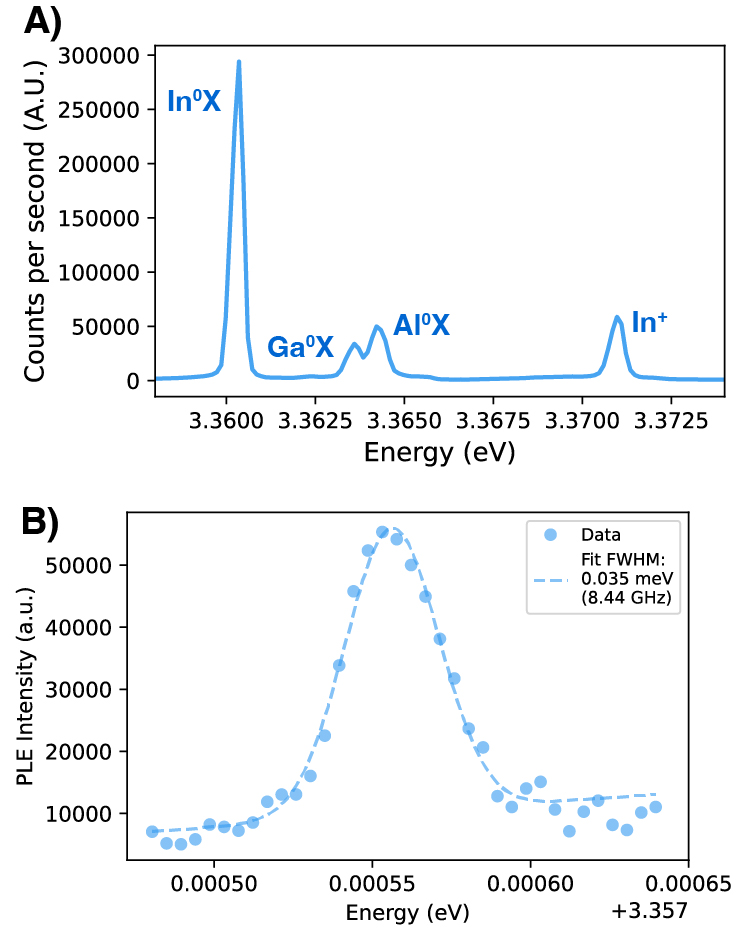}
    \caption{
    Continuous-wave excitation at 3.44\,eV, $\text{T} = 5.2$\,K.
    (a) Photoluminescence spectrum with $2$\,\textmu W of 360\,nm excitation. Emission lines corresponding to the Al, Ga, and In donor-bound exciton recombinations as well as the ionized In recombination are labeled accordingly.
    (b) Photoluminescence excitation plot utilizing side-excitation to collect the zero-phonon emission. The data are fit to a Voigt profile on a linearly sloped background, yielding a linewidth of $8.44$\,GHz. The excitation power is nominally $60$\,nW.
    }
    \label{fig:sample_char}
\end{figure}

We probe donor incorporation by measuring donor-bound-exciton (\DoX) recombination to donor-bound-electron (\Do) photoluminescence (PL) under $360$\,nm excitation. Figure~\ref{fig:sample_char}(a) shows a PL spectrum acquired from a region implanted with In donors, with the Al, Ga, and In donor-bound-exciton (\AlX, \GaX, and \InX) recombination lines labeled. The strong \InX\ emission indicates successful incorporation of In donors by implantation. We also observe weaker \AlX\ and \GaX\ emission. Because these donor species are present in the underlying substrate, and because the above-band excitation is not restricted to the implanted epilayer region, this background signal likely includes PL from below the epilayer surface rather than reflecting comparable Al or Ga concentrations within the implanted layer. We additionally observe emission associated with the ionized In donor (\InIon), which we attribute to partial compensation of the implanted donor ensemble by acceptors, either due to implantation or at the surface.

We further characterize the implanted In ensemble using photoluminescence excitation (PLE). In this measurement, we resonantly scan across the \transInX\ transition and collect the resulting \InX\ recombination, as shown in Fig.~\ref{fig:sample_char}(b). The PLE spectrum exhibits bright resonant emission from the implanted In donors and is well described by an ensemble linewidth of $8.44$\,GHz.

\section{MW Resonator}
\label{app:resonators}

\begin{figure*}[ht]
    \centering
    \includegraphics[width=0.99\linewidth]{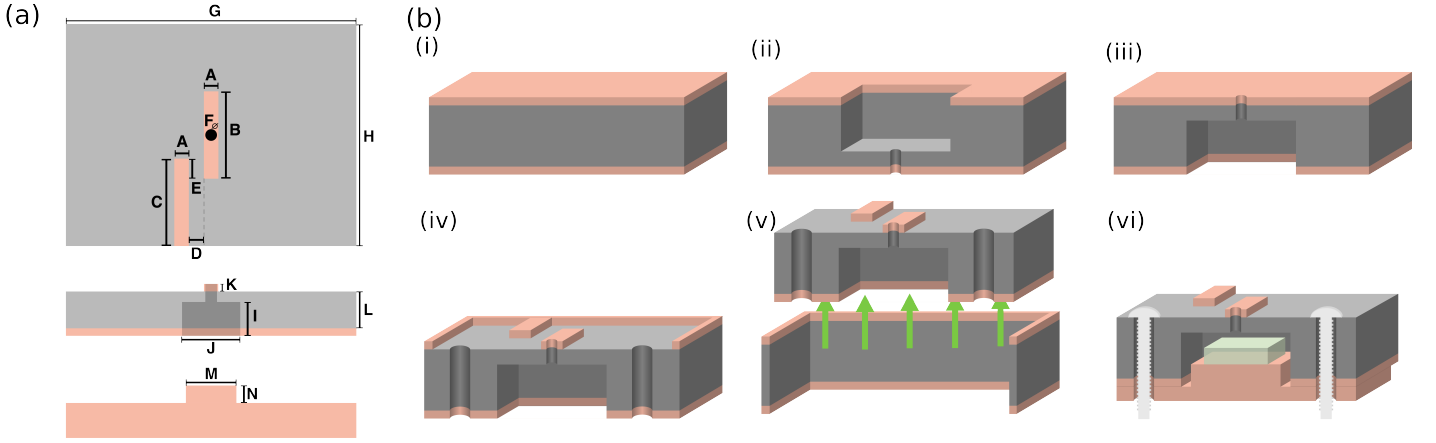}
    \caption{
    Detailed geometry and fabrication process for the gap-coupled microstrip resonator.
    (a) Dimensioned top-view and cross-sectional schematics of the resonator, sample pocket, and copper backplane. Dimensions labeled A--N are listed in Tab.~\ref{tab:resonator_dimensions}.
    (b) Fabrication sequence: (i) copper-clad Rogers RO3210 substrate, (ii) backside cutting of the sample pocket and optical access hole, (iii) alignment to the optical access hole for top-side patterning, (iv) patterning of the microstrip resonator, feedline, mounting holes, and device outline, (v) removal of resonator from substrate sheet, and (vi) completed resonator assembly with ZnO sample mounted to backplane and secured within nylon screws.
    }
    \label{fig:resFab}
\end{figure*}

Fig.~\ref{fig:resFab}(a) shows the key device dimensions used for the gap-coupled microstrip resonator. The top view defines the resonator trace, feedline, coupling gap, optical access hole, and overall chip dimensions, while the cross-sectional views show the substrate thickness, copper cladding, sample pocket, and copper backplane geometry. The labeled dimensions A--N are listed in Tab.~\ref{tab:resonator_dimensions}. These dimensions define the microwave geometry used in the field simulations and device fabrication.

The resonators were fabricated from Rogers RO3210 with $17.5\,\text{\textmu m}$ copper cladding on both sides and a substrate thickness of $1.27$\,mm. The resonator trace, feedline, optical access hole, and sample pocket were patterned using an LPKF Protolaser S3. The fabrication sequence is illustrated in Fig.~\ref{fig:resFab}. First, the copper-clad substrate was prepared as the starting material, Fig.~\ref{fig:resFab}(b-i). A rectangular sample pocket and optical access hole were then cut through the substrate from the ground-plane side, Fig.~\ref{fig:resFab}(b-ii). The substrate was then flipped, and the optical access hole was used as an alignment feature for patterning the top-side gap-coupled microstrip resonator and feedline, Fig.~\ref{fig:resFab}(b-iii). The top-side microwave traces and backplane mounting holes were patterned, after which the full resonator chip was cut from the substrate sheet, Fig.~\ref{fig:resFab}(b-iv). Finally, the device was removed from the sheet, an SMA connector was soldered to the feedline, and the ZnO sample mounted on a copper backplane was secured beneath the resonator using nylon screws, Fig.~\ref{fig:resFab}(b-v,b-vi).

\begin{table}[h]
    \centering
    \caption{Microwave resonator dimensions corresponding to the labels in Fig.~\ref{fig:resFab}.}
    \label{tab:resonator_dimensions}
    \begin{tabular}{c c}
        \hline
        \hline
        Label & Dimension \\
        \hline
        A & $1.00$\,mm \\
        B & $6.00$\,mm \\
        C & $6.075$\,mm \\
        D & $0.50$\,mm \\
        E & $0.95$\,mm \\
        F & $0.80$\,mm \\
        G & $20.00$\,mm \\
        H & $15.25$\,mm \\
        I & $0.95$\,mm \\
        J & $4.00$\,mm \\
        K & $0.0175$\,mm \\
        L & $1.27$\,mm \\
        M & $3.50$\,mm \\
        N & $0.40$\,mm \\
        \hline
    \end{tabular}
\end{table}

\bibliography{main.bib}

\end{document}